\documentclass[12pt,a4paper]{article}
\pdfoutput=1
\usepackage{amssymb}
\usepackage{amsmath}
\usepackage{amsfonts}
\usepackage{epsfig}
\usepackage{xcolor}
\usepackage[hidelinks]{hyperref}
\usepackage{cite}
\usepackage{calligra}
\usepackage[normalem]{ulem}
\usepackage{mathtools}
\usepackage{cite}
\usepackage{tikz-feynman}
\usepackage{tikz, pgf}
\usetikzlibrary{shapes.misc}
\usetikzlibrary{snakes}
\usetikzlibrary{decorations.pathmorphing}
\usetikzlibrary{shapes}
\usetikzlibrary{fit}
\usepackage{bm}
\outer\def\beginsection#1\par{\medbreak\bigskip
      \message{#1}\leftline{\bf#1}\nobreak\medskip
\vskip-\parskip
      \noindent}

\numberwithin{equation}{section}

\def\e{\epsilon}

\def\Ord{{\cal O}}

\def\be{\begin{equation}}
\def\ee{\end{equation}}
\def\ba{\begin{eqnarray}}

\newcommand{\ea}[1]{\begin{align} #1 \end{align}}

\def\nn{\nonumber \\}

\usepackage[titles]{tocloft}

\usepackage{titlesec}
\titleformat*{\section}{\large  \bfseries }
\titleformat*{\subsection}{\normalsize  \bfseries }

\begin{document}

\begin{titlepage}
\vskip 1.5cm
\begin{center}
{\Large \bf
Double-soft behavior of massless closed strings
interacting with any number of closed string tachyons
}

\vskip 1.0cm 
{\large 
Raffaele Marotta$^{a}$ and  Matin Mojaza$^{b}$
} \\[0.7cm] 
{\it $^a$  Istituto Nazionale di Fisica Nucleare, Sezione di Napoli, Complesso \\
 Universitario di Monte S. Angelo ed. 6, via Cintia, 80126, Napoli, Italy}\\[2mm]
 {\it $^b$  Max-Planck-Institut f\"ur Gravitationsphysik, \\
Albert-Einstein-Institut, Am M\"uhlenberg 1, 14476 Potsdam, Germany}
\end{center}
\begin{abstract}
We calculate the simultaneous double-soft limit of two massless closed strings scattering with any number of closed string tachyons to the subleading order at the tree level.
The limit factorizes the scattering amplitude into a double-soft factor multiplying the pure tachyon subamplitude,
suggesting a universal double-soft theorem for the massless closed string. 
We confirm an existing result for the double-soft graviton in an on-shell equivalent, but different form, while also establishing the double-soft factorization behavior of the string dilaton and of the Kalb-Ramond state, as well as the mixed graviton-dilaton case.
We also show that the simultaneous and consecutive double-soft theorems are 
consistent with each other.
We furthermore provide a complete field theory diagrammatic 
view on our result, which enables us in particular to establish
a four-point interaction vertex for two tachyons and two massless closed string states,
as well as the missing in field theory of three-point interaction of two massless closed string state and one tachyon.
\end{abstract}

{\let\thefootnote\relax
\footnotetext{
raffaele.marotta@na.infn.it, matin.mojaza@aei.mpg.de}
}

\vfill

\end{titlepage}


\setlength{\parskip}{5mm plus2mm minus2mm}

\tableofcontents

\vspace{5mm}
\clearpage

\setlength{\parskip}{3mm plus2mm minus2mm}

\section{Introduction}
\label{intro}
In this work we initiate the study of the double-soft behavior of massless closed strings
by considering their emission 
from $n$-point
closed string tachyon amplitudes in the bosonic string at the tree-level.
While the single-soft behavior is by now well-understood 
at the tree-level~\cite{Shapiro:1975cz,Ademollo:1975pf,Bianchi:2014gla,Schwab:2014sla,DiVecchia:2015oba,Guerrieri:2015eea,DiVecchia:2015srk,Bianchi:1512,DiVecchia:2016amo,DiVecchia:2016szw,Sen:2017xjn,DiVecchia:2017gfi,Higuchi:2018vyu,Marotta:2019cip} 
and to some extend also at the loop-level~\cite{Sen:2017nim,DiVecchia:2018dob,DiVecchia:2019kle}, the double-soft behavior have so far had little attention due to the large increase of complexity in the analysis (see, however, \cite{DiVecchia:2015bfa} for double-soft open strings).
The motivation has, however, not been lacking; 
It's been argued that for instance the double-soft behavior of massless states, 
in particular, of the graviton could shed new light on hidden symmetry of the gravitational S-matrix~\cite{Klose:2015xoa} in a way similar to the double-soft pion theorem, which reveals the underlying (hidden) coset algebra of spontaneously broken Quantum Chromodynamics~\cite{ArkaniHamed:2008gz}.
It has also been suggested~\cite{Wang:2015jna,Green:2019rhz} that the double-soft string dilaton theorem
could give novel constraints on the Type IIB supergravity action.
More fundamentally, since the origins of the dilaton soft theorem is not yet fully understood,
an extended understanding of its soft behavior could reveal key aspects to understand this problem.
Some hint on its origins could be deduced from the similarities existing between the single-soft theorems of the string dilaton and of the Nambu-Goldstone boson of spontaneously broken conformal symmetry. It has been shown that the soft-operators of both dilatons contain the generators of the conformal group, in particular dilations and special conformal transformations at leading and subleading order, respectively~\cite{DiVecchia:2015jaq,DiVecchia:2017uqn,Guerrieri:2017ujb}.
However, while the soft theorem of the Nambu-Goldstone dilaton
follows from  Ward identities associated with the broken generators of the conformal symmetry, a similar understanding of the soft behavior of the string dilaton
is still lacking.

Here we focus our attention on tree-level bosonic string amplitudes with two massless closed string states, representing gravitons, dilatons, and Kalb-Ramond B-fields, and an arbitrary number $n$ of closed string tachyons, which we denote as $M_{n+2}$. 
We will denote by $l$ and $q$ the momenta, and by $\varepsilon_{l}$ and $\varepsilon_{q}$ the polarization tensors, of the massless states, and with $k_1, \ldots, k_n$   the momenta of the tachyons.

These amplitudes can be expressed very compactly for arbitrary $n$
as integrals over the insertion points on a sphere of the vertex operators of each closed string. 
In this integral representation we are able to analyze generically $M_{n+2}$ for any $n$ in the kinematical region, where the momenta of the massless particles are simultaneously small; i.e. soft. 
 This is achieved by first rescaling with a small parameter $\tau$ the  two soft momenta,   $(l,\,q)\rightarrow (\tau l,\,\tau q)$,   and then by expanding the integrand  for small $\tau$, enabling us to calculate the integrals over the insertion points of the soft states up to the subleading order in $\tau$.
The result of these integrations
 is then found to be expressible in the form of an
operator acting on the pure $n$-point tachyon amplitude, $M_n$, specifically
\ea{
M_{n+2} (\tau q, \tau l, k_i)
= 
\varepsilon_{q,\mu \nu} \,\varepsilon_{l,\rho \sigma}
\left [ \frac{1}{\tau^2} S_{\rm W}^{\mu\nu\rho\sigma} (q,l, k_i)
\right.
&+ \frac{1}{\tau} S_{\rm E}^{\mu\nu\rho\sigma} (q,l,k_i)
\nn
& \left .
+ \frac{1}{\tau} \hat{S}_{J}^{\mu\nu\rho\sigma} (q,l,k_i)
\right ] M_n(k_i) + \Ord(\tau^0)
\label{0.0}
}
where the first two terms above are purely kinematical,
while the last term is operational (hence the hat), i.e. involving the angular momentum operator acting on the pure $n$-point tachyon amplitude.
{Their explicit expressions are specified after equations~\eqref{8.3}-\eqref{2.7}.}
The leading order term simply confirms the Weinberg theorem, but here includes also the contribution from the dilaton, while the subleading purely kinematical term $S_{\rm E}$ parametrizes, from a field theory perspective, contributions from double-soft emission directly from the same tachyon external state.

From this expression, the double-soft theorems for each of the physical states of the massless closed string, the graviton and dilaton respectively the Kalb-Ramond states, are obtained  upon symmetrization, respectively, antisymmetrization of the polarization indices of the massless legs.
For the graviton, the polarization indices have to additionally be transverse and traceless, while for the dilatons the polarization indices must be projected with a
transverse trace-tensor (see eq.~\eqref{dilatontensor}). 

The double-soft operator for external gravitons, given in \eqref{8.3} together with \eqref{9.2}-\eqref{9.4}, turns out to be in agreement with the known result in the literature~\cite{Chakrabarti:2017ltl,Chakrabarti:2017zmh}.
The double-soft theorems for two soft dilatons and for the mixed case of one graviton and one dilaton are obtained here, for the first time, and given 
in equations \eqref{DSD2} and \eqref{Mngd}, respectively.
Finally, the double-soft Kalb-Ramond emission theorem is expressed in \eqref{MnBB}-\eqref{M1BB}, where only the $S_{\rm E}$ term is non-vanishing.

Remarkably non of the soft operators above, obtained from string amplitudes,
contain string corrections (the stringy behavior is all contained in the factorized lower-point $n$-tachyon amplitude). Therefore in the field theory limit of the amplitudes, they remain unchanged providing relations among 
massive scalar amplitudes with and without two soft gravitons, dilatons, and Kalb-Ramond states\footnote{
The field theory effective action of tachyons found when keeping their mass 
$m^2=-\frac{4}{\alpha'}$ fixed, is that of massive $\phi^3$-interacting scalars~\cite{Scherk:71}}.
 We also remark that in the case of the dilaton, the soft-operator contains the generator of dilatations, as in the case of single-soft emission.

To outline the relation between the single-soft theorems and the double-soft theorems, found here, we derive from both theorems the consecutive soft emission of two soft massless closed string states; i.e. where the softness of the two states is ordered. This is detailed in the warm-up section~\ref{Single-soft}, where we revisit the single-soft theorem.
The two results agree upon symmetrization of the single-soft emission ordering, as should be expected. We stress that oppositely the double-soft theorem, being more general, cannot be derived from twice the application of the single-soft theorem.
On the other hand, the operator term $\hat{S}_J$ in \eqref{0.0} comes out fully from twice the application of the single-soft theorem, and we note that to the order considered, we do not see emergence of additional operators at the double-soft level.
In fact, we would expect, if at all, emergence of new operators first to appear at the order $\Ord(\tau^0)$, which, going beyond the scope of this work, motivates to extend our analysis to the subsubleading order, which we leave for future work.

To understand better all the additional purely kinematic terms appearing in 
$S_{\rm E}$, we perform in Sec.~\ref{diagrammatica} a field theory diagrammatic analysis of $M_{n+2}$, and show
the origin of each term in $S_{\rm E}$ from a field theory perspective.
This moreover enables us to predict from the double-soft theorem
the leading terms in the four-point Feynman vertex of two tachyons and two massless states. We furthermore consider in Sec.~\ref{Factorization} 
an alternative approach and show that $S_{\rm E}$, in fact, can be completely derived from on-shell factorization on the four-point subamplitude of two tachyons and two massless states, which we separately calculate in App.~\ref{App:4pt}.
The missing appearance of $\alpha'$ in the soft factors is then explained as a consequence of the property that the leading double-soft behavior of the four-point string amplitude is equivalent to its field theory limit $\alpha'\to 0$; a fact, that is also explained in Sec.~\ref{Factorization}. 
We also make use of the four-point amplitude to extract the contact terms, and thereby the full four-point on-shell vertex, confirming our prediction for it from the double-soft theorem. This analysis is provided in App.~\ref{APPB}.

To briefly recapitulate, the paper is organized as follows. 
In Sec.~\ref{Single-soft} we 
revisit
the single-soft theorems for massless closed string states and discuss the consecutive double-soft behavior of amplitudes with massless string states and  arbitrary numbers of massive scalar particles. 
In Sec.~\ref{simultaneous} we present the results of our analysis on the simultaneously double-soft limit from the perspective of the string amplitudes. In Sec.~\ref{Computational} and Sec.~\ref{gaugeinvariance} we give some technical details related to the analysis presented in Sec.~\ref{simultaneous}. In Sec.~\ref{diagrammatica} we discuss the diagrammatic analysis of the results obtained in the paper.
In Sec. \ref{Factorization} 
we use the on-shell factorization theorem to 
derive all the purely kinematical terms much more directly than in Sec.~\ref{diagrammatica}.

There are two appendices:  
App.~\ref{App:4pt} provides
the full string calculation of the four-point amplitude with two tachyons and two massless states, while App.~\ref{APPB} 
provides a field theory diagrammatic calculation of the same four-point amplitude, in its field theory limit, thereby establishing the four-point vertex of two tachyons and two massless closed string states.

\section{The single-soft case revisited}
\label{Single-soft}
Before we present our results,
it is useful and instructive to keep in mind the
single-soft scattering behavior of massless closed strings.
We thus give a brief summary thereof, 
and also discuss its application to two consecutively emitted soft massless closed strings, necessarily related to the more general result of this work.

\subsection{The single-soft theorem of a massless closed string}
The unified single-soft theorem for the states of the massless closed string; i.e the graviton, dilaton and Kalb-Ramond, was found in~\cite{DiVecchia:2017gfi} to read:
\ea{
M_{n+1}(k_i; q)= {\kappa_D}\, \e_{q,\mu}\bar{\e}_{q,\nu}
\sum_{i=1}^n& \Big[\frac{k_i^\mu k_i^\nu}{k_i \cdot q}-\frac{i}{2} \frac{k_i^\mu q_\rho \big(L_i+2\bar{S}_i\big)^{\nu\rho}}{k_i \cdot q}-   
\frac{i}{2} \frac{k_i^\nu q_\rho \big(L_i+2S_i\big)^{\mu\rho}}{k_i \cdot q} 
\nonumber\\
& 
+ \frac{i}{2}  \Big( S^{\mu \nu} -
 {\bar{S}}^{\mu \nu} \Big) \Big]M_n(k_i)
 + \Ord(q)
 \label{unified}
}
{where $\kappa_D$  is related to Newton's constant by 
$\kappa_D = \sqrt{8 \pi G_N^{(D)}}$}, with $D$ the number of spacetime dimensions,
and $\varepsilon_{q,\mu \nu} = \e_{q,\mu} \bar{\e}_{q,\nu}$ is the
polarization tensor of the massless state and $q$ is its momentum, which is soft compared to 
all the momenta $k_i$ of amplitude. 
$M_n$ is the lower-point amplitude without the soft state and universality of the soft theorem means that this can be any $n$-point amplitude. 
Finally the operators $L$, $S$ and $\bar{S}$ are defined as
\ea{
L_{i,\mu \nu} = i \left ( k_{i,\mu }\frac{\partial}{\partial k_{i}^{\nu}} - k_{i,\nu} \frac{\partial}{k_{i}^{\mu}} \right ) ,\,
S_{i\, \mu \rho} = i \left(\e_{i \mu}
\frac{\partial}{\partial \e_i^{\rho}} - \e_{i \rho}
\frac{\partial}{\partial \e_i^{\mu}} \right) , \, 
{\bar{S}}_{i\, \nu \sigma} = i
\left(\bar{\e}_{i \nu}
\frac{\partial}{\partial \bar{\e}_i^{\sigma}} - \bar{\e}_{i \sigma }
\frac{\partial}{\partial \bar{\e}_i^{\nu}} \right)  .
}
The full angular momentum operator is given by the combination
\begin{eqnarray}
J_{i,\mu\nu}=L_{i,\mu\nu}+S_{i,\mu\nu}+\bar{S}_{i,\mu\nu}=:L_{i,\mu\nu}+\mathcal{S}_{i,\mu\nu}\, ,\label{1.7}
\end{eqnarray}
where we also defined the total spin angular momentum operator $\mathcal{S}_i$.

The soft theorem in \eqref{unified} is a so-called subleading soft theorem; the factorization of the amplitude extends through subleading order in the soft momentum expansion.
It generically reproduces the soft behavior 
of the graviton, dilaton and B-field upon
symmetrization, respectively, antisymmetrization of the 
polarization tensor of the massless state. 
For the graviton one additionally imposes transversality, i.e. $q_\mu \varepsilon_{\rm g}^{\mu \nu} = 0$, while for the dilaton the polarization tensor is taken to be:
\ea{
\varepsilon_{{\rm d}}^{\mu \nu} = \frac{1}{\sqrt{D-2}} \left (\eta^{\mu \nu} - q^\mu \bar{q}^\nu - q^\nu \bar{q}^\mu \right ) \, , \qquad \bar{q}^2 = 0 \, , \ q\cdot \bar{q} = 1
\label{dilatontensor}
}
with $\bar{q}$ a reference null-vector.
In~\cite{DiVecchia:2017gfi} it was shown that the Kalb-Ramond $B$-field soft theorem does not extend to higher order in the soft expansion. However, it is well-known by now that the soft behavior of the graviton factorizes through subsubleading order at the tree-level, and this is also the case for the dilaton,
as shown in~\cite{DiVecchia:2016szw}. This behavior is for the dilaton, in contrast to the graviton, additionally universal and thus a true subsubleading soft theorem.
It is possible to write a unified subsubleading soft theorem for the graviton and dilaton, 
by considering only symmetric polarization of the soft massless closed state.
The first two orders can be immediately derived from \eqref{unified}, 
while the subsubleading operator was given for all string theories in~\cite{DiVecchia:2016szw}.
The full expression takes the form
\ea{
M_{n+1}  &= \kappa_D \, \e_{q,\mu}\e_{q, \nu} 
\sum_{i=1} 
\left [\frac{k_i^\mu k_i^\nu}{k_i \cdot q}
 - i \frac{q_\rho 
k_i^\nu  J_i^{\mu \rho}}{k_i \cdot q}
 -
\frac{q_\rho q_\sigma :J_i^{\mu \rho}  J_i^{\nu \sigma}:}{2k_i \cdot q}
-
\frac{\alpha'}{2} \frac{q_\rho q_\sigma
\hat{S}_{i, \rm t.d.}^{\mu \rho, \nu \sigma}
}{k_i \cdot q} 
\right ]M_n + \Ord(q^2)
 }
 where $::$ means normal ordering of the double-derivatives such that they all act to the right,
 and the last term, clearly theory dependent (t.d.) due to the string parameter $\alpha'$,
 is given by:
\ea{ 
\hat{S}_{i, \rm t.d.}^{\mu \rho, \nu \sigma}=
 \left (k_i^\mu \eta^\rho_\alpha - k_i^\rho \eta^\mu_\alpha\right )
  \left (k_i^\nu \eta^\sigma_\beta - k_i^\sigma \eta^\nu_\beta\right )
\Pi_{i, {\rm t.d.}}^{\alpha \beta}
}
with
\ea{
\Pi_{i, {\rm t.d.}}^{\alpha \beta} =
\left\{
\begin{array}{ll}
	\e_{i}^\alpha \frac{\partial}{\partial \e_{i,\beta}}
+\bar{\e}_{i}^\alpha \frac{\partial}{\partial \bar{\e}_{i,\beta}}
& \text{ bosonic string} \\
	\e_{i}^\alpha \frac{\partial}{\partial \e_{i,\beta}}
& \text{ heterotic string} \\
0 & \text{ superstring} 
\end{array}
\right.
\label{Pi}
}
Thus, the graviton in superstring theory has the same soft factorization behavior as the field theory graviton of ordinary Einstein-Hilbert gravity. 
These theory dependent contributions are directly related to the difference in their low-energy actions at the leading $\alpha'$ level, where in bosonic and heterotic string enters
a $R^2 \phi$ term in the action (where $R^2$ should be understood as the Gauss-Bonnet operator and $\phi$ is the dilaton field), which does not appear in superstrings due to supersymmetry.
{It turns out, however that}
 the theory dependent term vanishes (on shell) in all theories, when contracted with the dilaton projection tensor. 
This property ensures that the dilaton soft behavior is universal, and 
takes the curious form when use of \eqref{dilatontensor}, momentum conservation and on-shell conditions, is made:
\ea{
M_{n+\phi} = 
\frac{\kappa_D}{ \sqrt{D-2} }  &\left[ - \sum_{i=1}^n \frac{m_i^2}{k_i q} {\rm e}^{q \partial_{k_i}}  + 2 - \sum_{i=1}^n\hat{D}_i
+  q_\mu \sum_{i=1}^n {\hat{K}}_{i}^{ \mu}
\right.
\nonumber \\
&\left . \quad 
+
 \sum_{i=1}^n
\frac{q_\rho q_\sigma}{ 2 k_i q}
\left(
\mathcal{S}_{i}^{ \rho \mu}\eta_{\mu \nu} \mathcal{S}_{i}^{ \nu \sigma} + D
\Pi_{i, \rm bos.}^{\rho \sigma} \right) 
   \right]  M_n   + \Ord(q^2)
   \label{singlesoftdilaton}
}
where
\ea{
\hat{D}_i = k_{i} \cdot  
\frac{\partial}{\partial k_{i}}  \, , \qquad 
 \hat{K}_{i}^{\mu} = \frac{1}{2} k_{i}^{ \mu} \frac{\partial^2}{\partial
k_{i\nu} \partial k_i^\nu} 
-k_{i}^{\rho} \frac{\partial^2}{ \partial k_{i}^{\rho} \partial
k_{i\mu}} 
-i \mathcal{S}_i^{\mu\rho}\frac{\partial}{\partial k_i^\rho}
 \, ,
}
are the generators of the space-time dilatations and special conformal transformations. 
the subscript bos. on $\Pi_i^{\rho \sigma}$ means that it is the operator identically to \eqref{Pi} in the case of the bosonic string, but here enters universally for all string theories.
The universality of this soft theorem persists even at the loop level, 
albeit one has to rewrite the dilatation operator in terms rescaling operators, see~\cite{DiVecchia:2019kle}.

\subsection{The consecutive soft theorem of two massless closed strings}
\label{Consecutive-soft}
Before we present the double-soft results, 
it is useful to first consider what information the consecutive emission
of two soft massless states provides.
To keep things simple, but still giving the idea, we 
restrict to the consecutive double-soft emission of two dilatons
from an $n$-point closed string tachyon amplitude.
For comparison with the simultaneous soft emission, 
where the two soft dilatons are not distinguished,
we symmetrize the consecutive order of soft limits.
To subleading order we have:
\begin{align}
\frac{1}{2}&\left\{	\lim_{l\to 0} , \lim_{q\to 0} \right \}M_n^{dd} (q,l; k_i)
= \frac{1}{2} \frac{\kappa_D^2}{D-2} \left [ - \sum_{i=1}^n \frac{m_i^2}{q\cdot k_i} \left (1 + q \cdot \partial_{k_i} \right ) + 2 - \sum_{i=1}^n k_i \cdot \partial_{k_i} - l \cdot \partial_l \right ]
\nonumber \\
&\times  \left [ - \sum_{j=1}^n \frac{m_j^2}{l\cdot k_j} \left (1 + l \cdot \partial_{k_j} \right ) + 2 - \sum_{j=1}^n k_j \cdot \partial_{k_j}\right ] M_n (k_i) + (q \leftrightarrow l) + \Ord(q^0,l^0)
\end{align}
Expanding the two soft brackets (acting also with the first operator on the second) and rearranging one finds
\begin{align}
\frac{1}{2}&\left\{	\lim_{l\to 0} , \lim_{q\to 0} \right \}M_n^{dd} (q,l; k_i)
= \frac{\kappa_D^2}{D-2} \left \{ 
\sum_{i,j=1}^n\frac{m_i^2\,m_j^2}{ k_iq\, k_jl} \left( 1+  q\frac{\partial}{\partial k_i}+ l\frac{\partial}{\partial k_j}\right) 
\right . \nonumber \\
&\left .
+
\sum_{i=1}^n\left(\frac{m_i^2}{k_iq}+\frac{m_i^2}{ k_il}\right)\sum_{j=1}^n k_j\frac{\partial }{\partial k_j}
- 3 \sum_{i=1}^n\left(\frac{m_i^2}{ k_iq}+\frac{m_i^2}{ k_il}\right)\right \}M_n(k_i) 
+ \Ord(q^0, l^0)
\label{9.29}
\end{align}
The first two series of terms follow immediately from the single-soft theorem.
The last series of terms is, however, not obvious. In particular
the prefactor 3, instead of just 2, origins, as we will see, 
from the contribution of the four-point contact term of two tachyons and two dilatons.
Here the extra factor appears as a consequence of the dilation operator in the soft theorem.
This shows highly non-trivially that the single-soft dilaton theorem stores information about higher-point interactions, and these are encoded in the conformal operators.

The more general consecutive soft emission of two (or more) massless closed string
can be obtained in the same way directly from the unified single-soft theorem
\eqref{unified}.
However, this will be much easier to obtain, once we have provided
the simultaneous double-soft theorem.

\section{The simultaneous double-soft theorem of massless closed strings}
\label{simultaneous}
We now present a summary of our main results.
In this work, we restrict the analysis to the double-soft emission
of massless closed strings from $n$ closed string tachyons in the bosonic string.
The full string amplitude can be written in closed form for any $n$ tachyons as follows:
\begin{eqnarray}
&&M_{2;n}(q, l, \{k_i\}) =N_0^{n+2}\,C_0 \int \prod_{i=1}^n \frac{d^2z_i\, d^2 z\,d^2 w}{dV_{abc}}\int d\theta d\varphi \prod_{i=1}^N d\theta_i \,\int d\bar{\theta} d\bar{\varphi} \prod_{i=1}^N d\bar{\theta}_i\,\nonumber\\
&&\times \Bigg[{e^{\frac{\alpha'}{4} ql G(z,w)}}\,e^{(\theta\e_q)(\varphi\e_l)\partial_z\partial_wG(z,w)+\sqrt{\frac{\alpha'}{2}}l(\theta\e_q)\partial_zG(z,w)+\sqrt{\frac{\alpha'}{2}}q(\varphi\e_l)\partial_wG(z,w)}\nonumber\\
&&\times \prod_{i=1}^n {e^{\frac{\alpha'}{4} k_iq\,G(z,z_i)}}\times \prod_{i=1}^n e^{\sqrt{\frac{\alpha'}{2}} k_i(\theta\e_q) \partial_z G(z,z_i)} \times \prod_{i=1}^n {e^{\frac{\alpha'}{4} k_il G(w,z_i) }}\times \prod_{i=1}^n e^{\sqrt{\frac{\alpha'}{2}} k_i(\varphi\e_l) \partial_w G(w,z_i)}  \nonumber\\
&&\times \prod_{i<j}e^{\frac{\alpha'}{4} k_ik_j G(z_i,z_j)}\Bigg] 
\times  \text{a.h.}\label{2.1}
\end{eqnarray}
where $q,l$ are the momenta of the massless closed strings, $\e_q, \e_l$ are their holomorphic polarization vector, and $k_i$ are the momenta of the tachyons.
$\theta$ and $\varphi$ are Grassmann variables and a.h. stands for multiplication with the corresponding antiholomorphic part. $G(z,w) = \log |z-w|^2$ is the two-point Green function on the Riemann sphere, however, for our purpose, and for possible extension to multiloops, 
it is useful keep the expression in terms of the Green functions.
The prefactors are given by:
\ea{
C_0 = 
\left(\frac{8\pi}{\alpha'}\right)\left(\frac{2\pi}{\kappa_D}\right)^2
\, , \qquad
N_0 =
\left(\frac{\kappa_D}{2\pi}\right) \, ,\qquad 
N_0^{n+2} C_0 = \left(\frac{8\pi}{\alpha'}\right) \left(\frac{\kappa_D}{2\pi}\right)^n
}
After performing integration over the Grassmann variables,
the expression can be brought into the following form:
\begin{eqnarray}
&&M_{n+2}(\{k_i\}; q, l) =M_n(\{k_i\})\star\e_q^\mu \,\e_l^\nu\,\bar{\e}_q^\rho\,\bar{\e}_l^\sigma  N_0^2\int d^2z\,d^2w \,e^{\frac{\alpha'}{2} ql G(z,w)}\, \prod_{i=1}^n e^{\frac{\alpha'}{2} k_iq\,G(z,z_i)}\,\prod_{i=1}^n e^{\frac{\alpha'}{2} k_il G(w,z_i) }\nonumber\\
&&\times \left[\eta_{\mu\nu} \,\partial_z\partial_w G(z,w)+\frac{\alpha'}{2} \sum_{i,j=1}^nk_{i\mu}\,k_{j\nu} \partial_zG(z,z_i)\,\partial_wG(w,z_j)  +\frac{\alpha'}{2} \sum_{j=1}^n l_\mu k_{j\nu} \partial_z G(z,w) \partial_wG(w,z_j)\right.\nonumber\\
&&\left.+\frac{\alpha'}{2} \sum_{i=1}^n q_\nu k_{i\mu} \partial_z G(z,z_i)\,\partial_wG(z,w)+\frac{\alpha'}{2} l_\mu q_\nu\partial_z G(z,w)\partial_w G(z,w)\right] \nonumber\\
&&
\times\Bigg[\eta_{\rho\sigma} \,\partial_{\bar{z}}\partial_{\bar{w}} G({z},{w})
+\frac{\alpha'}{2} 
\sum_{i,j=1}^n k_{i\rho}\,k_{j\sigma} \partial_{\bar{z}}G({z},{z}_i)\,\partial_{\bar{w}}G({w},{z}_j)+\frac{\alpha'}{2} \sum_{j=1}^n l_\rho k_{j\sigma} \partial_{\bar{z}} G({z},{w}) \partial_{\bar{w}}G({w},{z}_j)\nonumber\\
&& \left.+\frac{\alpha'}{2} \sum_{i=1}^n q_\sigma k_{i\rho} \partial_{\bar{z}} G({z},{z}_i)\,\partial_{\bar{w}}G({z},{w})+\frac{\alpha'}{2} 
l_\rho q_\sigma\partial_{\bar{z}} G({z},{w})\,\partial_{\bar{w}} G({z},{w})
\right]
\label{Mn+2}
\end{eqnarray}
where $M_n(\{k_i\})$ is the $n$-point tachyon amplitude and $\star$ denotes a convolution 
of integrals. Our aim is to calculate the integration
over the massless closed string vertex operator positions, parametrized by $z$ and $w$,
up to subleading order in $q$ and $l$; more specifically, 
let $q^\mu \to \tau q^\mu$ and $l^\mu \to \tau l^\mu$, we will compute through order $\tau^{-1}$. Then we will seek an operator, which when acting on the pure tachyon amplitude in integral form, reproduces our calculation. This will establish a double-soft theorem, if such an operator exists.

The result of our calculation of the $z$ and $w$ integrals
is not in it self interesting. 
But from that calculation we are able to establish the following result:
\ea{
 M_{n+2}^{\mu\nu,\rho\sigma}&(\{k_i\}; \tau q, \tau l)
=\kappa_D^2\, 
\varepsilon_{q,\mu \nu} \,\varepsilon_{l,\rho \sigma}
\left\{
\frac{1}{\tau^{2}} \sum_{i,j=1}^n \frac{k_i^\mu \,k_i^\nu}{ k_iq}\,\frac{k_j^\rho\,k_j^\sigma}{ k_jl}
\right .
\nonumber \\
& \left .
+\frac{1}{\tau}\sum_{i=1}^n \frac{1}{ k_i(q+l)}\left[\frac{M_1^{\mu\nu\rho\sigma}}{ql}+\frac{M_2^{\mu\nu\rho\sigma}}{(k_iq)\,(k_il)}+ M_3^{\mu\nu\rho\sigma} \right]
\right.\nonumber\\
&\left.
-\frac{i}{\tau} \sum_{i=1}^n \frac{k_i^\rho\, k_i^\sigma}{k_il} \,\sum_{j=1}^n\frac{q_\tau k_j^\mu\, J_j^{\nu\tau}}{k_jq} 
-\frac{i}{\tau} \sum_{i=1}^n \frac{k_i^\mu k_i^\nu }{k_iq} \,\sum_{j=1}^n \frac{l_\tau k_j^\rho J_j^{\sigma\tau}}{k_jl}\right\}M_n(\{k_i\})
+ \Ord(\tau^0)
\label{8.3}
}
The coefficients $M_{1,2,3}= \varepsilon_q^{\mu \nu} \,\varepsilon_l^{\rho \sigma} M_{1,2,3}^{\mu\nu\rho\sigma}$ are given by:
\ea{
M_1 = &-(\varepsilon_q^{\mu\nu} \, \varepsilon_{l\mu\nu})\,(q k_i) \,(l k_i)+(l\varepsilon_ql)\,(k_i\varepsilon_lk_i)+(k_i\varepsilon_qk_i)\,(q\varepsilon_lq)-2(l\varepsilon_qk_i)\,(k_i\varepsilon_lq)\nonumber\\
&+[(l \varepsilon_q^t\, \varepsilon_l k_{i})+(l\varepsilon_q\varepsilon_l^tk_i)] (k_iq)+[
 (k_i\varepsilon_q^t\,\varepsilon_l q)+(k_i\varepsilon_q\varepsilon_l^tq)](k_il)\label{2.5}
\\[2mm]
M_2 =&
\, (k_i\varepsilon_qk_i)[(k_i\varepsilon_lq)+(q\varepsilon_lk_i)](k_il)
+ [(l\varepsilon_qk_i)+(k_i\varepsilon_ql)]\,(k_i\varepsilon_lk_i)(qk_i)
\nonumber\\
&
-(k_i\varepsilon_qk_i)\,(k_i\varepsilon_lk_i)(ql)
\label{2.6}
\\[2mm]
M_3 =&-[(k_{i} \varepsilon_q^t\,\varepsilon_{l} k_i)+(k_i\varepsilon_q\varepsilon_l^tk_i)]
\label{2.7}
}
{The operators $S_W^{\mu\nu;\rho \sigma}$, $S_E^{\mu\nu;\rho \sigma}$ and $S_J^{\mu\nu;\rho \sigma}$
introduced in the introduction, Eq.~\eqref{0.0}, are given by the first, second and third line of \eqref{8.3}, respectively.}
 
Since the lower-point amplitude $M_n$ is a purely tachyonic amplitude,
the action of $J^{\mu \nu}$ is here simply that of $L^{\mu \nu}$, 
but for a generalization of this soft theorem, one should expect the full $J^{\mu \nu}$ to appear, why we have kept it in this form here (the $\mathcal{S}^{\mu \nu}$-part here simply annihilates the lower-point amplitude, so its addition is here indifferent).

We emphasize that \eqref{8.3} encodes the result of the explicit computation of 
\eqref{Mn+2}. It specifically shows that the double-soft emission of 
massless closed strings obeys a double-soft theorem through subleading order.
{It will be checked in Sec.~\ref{gaugeinvariance}} 
that \eqref{8.3} is on-shell gauge invariant, as it should be.

We finally notice that \eqref{8.3} does not contain any explicit $\alpha'$-terms.

In the following subsections we obtain from \eqref{8.3} the specific double-soft theorem associated to each of the massless string states by contracting it with the corresponding polarization tensors. 
We notice first, however, that the expression must be different from zero only if the soft states are either both symmetrically polarized (gravitons/dilatons) or both Kalb-Ramond (antisymmetric) states. This is a consequence of the world-sheet parity symmetry of the bosonic string, according to which amplitudes odd under the exchange of the left and right sectors of the closed string vanish. It is for later use, i.e. when checking gauge invariance in Sec.~\ref{gaugeinvariance}, useful to rewrite \eqref{2.5} in 
a way that takes this symmetry into account, namely
\ea{
M_1 =& -(\varepsilon_q^{\mu\nu} \, \varepsilon_{l\mu\nu})\,(q k_i) \,(l k_i)+(l\varepsilon_ql)\,(k_i\varepsilon_lk_i)+(k_i\varepsilon_qk_i)\,(q\varepsilon_lq)-(l\varepsilon_qk_i)\,(k_i\varepsilon_lq)\nonumber\\
&-(k_i\varepsilon_q l)\,(q\varepsilon_lk_i)+[(l \varepsilon_q^t\, \varepsilon_l k_{i})+[(l\varepsilon_q\varepsilon_l^tk_i)] (k_iq)+[
 (k_i\varepsilon_q^t\,\varepsilon_l q)+(k_i\varepsilon_q\varepsilon_l^tq)](k_il)\label{4.2}
}
where only the term with a factor of 2 in \eqref{2.5} was rewritten.

\subsection{Two symmetrically polarized soft states (gravitons and dilatons)}
In the case where $\varepsilon_q^{\mu\nu}$ and $\varepsilon_l^{\mu\nu}$ are both symmetric,
the coefficients $M_{1,2,3}$ simplify to:
\ea{
 M_1=& -(\varepsilon_q^{\mu \nu} \varepsilon_{l,\mu \nu})\,(q k_i) \,(l k_i)+(l\varepsilon_ql)\,(k_i\varepsilon_lk_i)+(k_i\varepsilon_qk_i)\,(q\varepsilon_lq)\nonumber\\
 &
 -2(l\varepsilon_qk_i)\,(q\varepsilon_lk_i)
 +2(l\varepsilon_q \varepsilon_l k_i) (k_iq)+2
 (k_i\varepsilon_q\varepsilon_l q)(k_il)
 \label{9.2}
\\[2mm]
M_2 =&
2 (k_i\varepsilon_qk_i)\,(k_i\varepsilon_lq)(k_il)
+ 2(l\varepsilon_qk_i)\,(k_i\varepsilon_lk_i)(qk_i)
-(k_i\varepsilon_qk_i)\,(k_i\varepsilon_lk_i)(ql)
\label{9.3}
\\[2mm]
M_3 =&-2 (k_i \varepsilon_q\varepsilon_l k_i)
\label{9.4}
}

For two soft gravitons, our result has to agree with the already established field theory
double-soft graviton expressions~\cite{Chakrabarti:2017ltl}.
Eq.~\eqref{8.3} is, however, different from the result of~\cite{Chakrabarti:2017ltl} for the graviton. The difference is in the $M_1$-term,
but we can show that the two expressions are equivalent 
as a consequence of momentum conservation;
i.e. we can write the terms in question as:
\ea{
\sum_{i=1}^n\frac{1}{ k_i(l+q)\,ql}
&
\left[  -(\varepsilon_q^{\mu \nu} \varepsilon_{l,\mu \nu})\,(q k_i) \,(l k_i)+(l\varepsilon_ql)\,(k_i\varepsilon_lk_i)+(k_i\varepsilon_qk_i)\,(q\varepsilon_lq)
-2(l\varepsilon_qk_i)\,(q\varepsilon_lk_i)
\right.\nonumber\\
 &
 +(l\varepsilon_q \varepsilon_l k_i) (k_i(q-l))+
 (k_i\varepsilon_q\varepsilon_l q)(k_i(l-q))
 \nonumber\\
 &\left.+(l\varepsilon_q \varepsilon_l k_i) (k_i(q+l))+ (k_i\varepsilon_q\varepsilon_l q) (k_i(q+l))\right]
 \label{Sen}
}
In the last line the numerator cancels the pole in $k_i(q+l)$ and, due to the momentum conservation,  the sum over the hard particles yields:
\ea{
\sum_{i=1}^n
\left [(l\varepsilon_q \varepsilon_l k_i) + (k_i\varepsilon_q\varepsilon_l q) \right ]
= - 2(l\varepsilon_q \varepsilon_l q)
}
which can be neglected to the order in the soft expansion we are considering.
The remainder of \eqref{Sen} then agrees with~\cite{Chakrabarti:2017ltl}.

For the simultaneous soft emission of both a graviton and a dilaton,
we can specify the result further by using the dilaton projection tensor~\eqref{dilatontensor} and that the graviton polarization tensor is traceless,
thereby getting
\ea{
M_{n+g+d}= &-\frac{\kappa_D^2}{\sqrt{D-2}}\Bigg[
\frac{1}{\tau^2} \sum_{j=1}^n \frac{k_j\varepsilon_lk_j}{ k_jl} \sum_{i=1}^n\frac{m_i^2}{k_iq} \left(1+\tau\,q\frac{\partial}{\partial k_i}\right)
+\frac{1}{\tau}\sum_{j=1}^n \frac{k_j\varepsilon_lk_j}{k_jl} \sum_{i=1}^n k_i\frac{\partial}{\partial k_i}
\nonumber \\
&
+\frac{1}{\tau}\sum_{i=1}^n \frac{m_i^2}{k_i(q+l)} \left( \frac{q\varepsilon_lq}{ql} +2\frac{k_i\varepsilon_lq}{k_iq} -\frac{(ql)(k_i\varepsilon_lk_i)}{(k_il)(k_iq)} \right)\nonumber\\
&-\frac{i}{\tau}\sum_{i=1}^n\frac{m_i^2}{ k_iq} \sum_{j=1}^n\frac{\varepsilon_{l\mu\nu} k_j^\mu l_\rho  J_j^{\nu\rho}}{k_jl}\Bigg]M_n + \Ord(\tau^0)
\label{Mngd}
}

Finally we can specify to the double-soft dilaton case, getting
(we set $k_i^2 = -m_i^2$):
\ea{
M_{n+d+d} =
\frac{\kappa_D^2}{D-2 } 
\Bigg [&
\frac{1}{\tau^2} \sum_{i,j=1}^n\frac{m_i^2\,m_j^2}{k_iq\, k_jl} 
\left( 1+ \tau q\frac{\partial}{\partial k_i}+\tau l\frac{\partial}{\partial k_j}\right)
-\frac{1}{\tau} 
\sum_{i=1}^n \frac{m_i^4}{(k_il)(k_iq) } \frac{(ql)}{k_i(q+l)}
\nonumber\\
&
- \frac{D-2}{\tau} \sum_{i=1}^n \frac{(qk_i)(lk_i)}{ k_i(q+l)\,(ql)} 
-\frac{1}{\tau} 
\sum_{i=1}^n \frac{2 m_i^2}{k_i(q+l)} 
\nonumber\\
&
-\frac{1}{\tau}\sum_{i=1}^n\left(\frac{m_i^2}{ k_iq}+\frac{m_i^2}{ k_il}\right)
\left ( 2-\sum_{j=1}^n k_j\frac{\partial }{\partial k_j} \right )
\Bigg]M_n  + \Ord(\tau^0)
 \label{DSD2}
}
Since these results apply to $n$ hard states being closed string tachyons,
the masses should be taken equal to $m_i^2=-4/\alpha'$, 
however, we expect that this result applies, at least to some extend, more generally to the interaction with other massive bosons.
As will be detailed later,
all terms with a double-pole in the soft limit
can be understood as double-soft 
emissions directly from
external tachyon lines through three-point interactions.
The last line, with only a simple pole structure, 
can be directly understood from the single-soft theorem \eqref{singlesoftdilaton}.
Finally there remains one additional term with a simple pole;
this is also an external line emission contribution, however, its structure
reveals that it must come from a four-point contact interaction where
two dilatons are emitted from the same point of an external line.
In other words, the double-soft theorem immediately give us the on-shell contact interaction of two tachyons with two dilatons.

The consecutive soft theorems can be immediately derived from these results.
As an example, let us consider again
the consecutive emission of two soft dilatons and 
and compare with \eqref{9.29}.
Specializing \eqref{DSD2} to the symmetrized consecutive soft theorem, we get:
\begin{align}
\frac{1}{2}\left\{	\lim_{l\to 0} , \lim_{q\to 0} \right \}&M_{n+d+d}
=  \frac{\kappa_D^2}{D-2} \Bigg [
\sum_{i,j=1}^n\frac{m_i^2\,m_j^2}{ k_iq\, k_jl} \left( 1+  q\frac{\partial}{\partial k_i}+ l\frac{\partial}{\partial k_j}\right) 
\nonumber \\
&
+
\sum_{i=1}^n\left(\frac{m_i^2}{k_iq}+\frac{m_i^2}{ k_il}\right)\sum_{j=1}^n k_j\frac{\partial }{\partial k_j}
-3
\sum_{i=1}^n \left(\frac{m_i^2}{ k_iq}+\frac{m_i^2}{ k_il}\right)
\nonumber \\
&
-
{(D-2)\sum_{i=1}^n \frac{k_i (q + l)}{2(ql)}}
-
\sum_{i=1}^n \frac{m_i^4  (ql)}{2(k_il) (k_i q)} 
\left ( \frac{1}{k_i q}+\frac{1}{k_i l} \right )
\Bigg ]M_n  + \Ord(q^0,l^0)
\end{align}
We observe that the first two lines are exactly the same as what we found
from using the single-soft theorem consecutively in \eqref{9.29}.
But now we understand the factor $3$ discussed below \eqref{9.29}; it is the additional contribution from the four-point contact term of \eqref{DSD2} that here ensures the factor of 3. Thus the single-soft theorem, used to derived \eqref{9.29}, rather amazingly contains the information about the four-point contact term.

\subsection{Two antisymmetrically polarized (Kalb-Ramond) soft states}

Since most terms of \eqref{8.3} are symmetric in the polarization indices of at least one of the soft massless states, the double-soft theorem \eqref{8.3} simplifies radically when taking the external states to be antisymmetrically polarized.
We specifically find that the soft theorem of two Kalb-Ramond B-fields read:
\begin{eqnarray}
&&M_{n+B+B}=\kappa_D^2  \sum_{i=1}^n \frac{1}{\tau k_i(q+l)}\left[\frac{M_1^{(BB)}}{ql}+2 (k_{i} \varepsilon_q\,\varepsilon_l\, k_i) \right]M_{n} + \Ord(\tau^0)
\label{MnBB}
\end{eqnarray}
with
\ea{
M_1^{(BB)}= &&= (\varepsilon_q^{\mu\nu} \, \varepsilon_{l,\,\nu\mu})\,(q k_i) \,(l k_i)-2(l \varepsilon_q k_i)\,(k_i \varepsilon_l q)
 -2(l \varepsilon_{q}\, \varepsilon_l k_{i}) (k_iq)-2
 (k_i \varepsilon_{q}\,\varepsilon_l q)(k_il)
 \label{M1BB}
}
We recall that the amplitude of a single B-field interacting with $n$-tachyons vanishes, and so does its soft theorem; i.e. the B-field single-soft operator annihilates the pure tachyon amplitude. Now, instead with two soft B-fields, the soft behavior is non-vanishing, and given by the above expression.

\section{Computational details}
\label{Computational}
In this section we give some detail about the derivation of the simultaneous double soft limit results presented in sec. \ref{simultaneous}.  The advantage of using string amplitudes to find such relations, which to leading order in the string slope are field theory identities, is due to their peculiarity to be very compact expressions   containing few diagrams, only one in the case of oriented closed theories,  at each order or the perturbative expansion. This is visible in Eq. \eqref{2.1} where the  amplitude, giving the interaction among two massless states and $n$-tachyons, is a multiple integral on the complex Koba-Nielsen variables parametrizing the insertion of the string vertices on the complex plane  $\mathbb{C}P^1$.
The amplitude turns out to be a convolution integral among the $n$-tachyon amplitude and two extra integrals that collect all the dependence on the complex variables, $w$ and $z$, associated to the two massless vertices carrying soft momenta $q$ and $l$. The soft limit is obtained by performing, for small values of $l$ and $q$,  the integration on these two complex variables. The integrals to be evaluated for low momenta are all collected by the  following general expression:
\begin{eqnarray}
I^{n m a_i b_j}_{\bar{n}\bar{m} \bar{a}_i\bar{b}_j}=&&\!\!\int d^2 z d^2 w
e^{\frac{\alpha'}{2}\tau^2 ql G(z,w)}\prod_{i=1}^ne^{\frac{\alpha'}{2}\tau k_iq G(z,z_i)}\prod_{i=1}^ne^{\frac{\alpha'}{2} \tau k_i lG(w,z_i)}\,[\partial_z G(z,w)]^n\,[\partial_wG(z,w)]^m\nonumber\\
\times &&\!\![\partial_zG(z,z_i)]^{a_i}\,[\partial_wG(z,z_j)]^{b_j}\, [\partial_{\bar{z}}G(z,w)]^{\bar{n}}\,[\partial_{\bar{w}} G(z,w)]^{\bar{m}}\,[\partial_{\bar{z}}G(z,z_i)]^{\bar{a}_i}\,[\partial_{\bar{w}}G(z,z_j)]^{\bar{b}_j}\nonumber\\
&&\label{3.1}
\end{eqnarray}
 with $(n,\,m,\,a,\,b,\,\bar{n},\,\bar{m},\bar{a},\bar{b})=0,\,1$. 
 Eq. \eqref{Mn+2} contains also terms with the double derivatives  of the Green function, i.e. $\partial_z\partial_wG(z,w)$ and analogous  anti-holomorphic expressions, which may not seem described by \eqref{3.1}. However, due to the identity 
 $\partial_z\partial_wG(z,w)=-\partial_wG(z,w)\partial_zG(z,w)$, valid for the tree level Green function $G(z,w) = \log |z-w|^2$, the general expression \eqref{3.1} 
 therefore does in fact describe those terms as well.

There are six different typologies of such integrals depending of the degrees of divergence for $w\sim z$.
The  one with highest pole is\footnote{$I^{1100}_{1100}$  is infrared divergent  for $w\simeq z$  in the kinematic region  $\tau^2(q+l)^2\leq \frac{4}{\alpha'}$, with $-4/\alpha'$ being the tachyon mass. We compute this integral in the kinematic region where it is well defined and then we analytically extend it in the soft region $\tau\simeq 0$, cf.~\cite{Witten:2013pra}.}:
\begin{eqnarray}
I^{1100}_{1100}=&& \int d^2 z d^2 w\,
e^{\frac{\alpha'}{2}\tau^2 ql G(z,w)}\prod_{i=1}^ne^{\frac{\alpha'}{2}\tau k_iq G(z,z_i)}\prod_{i=1}^ne^{\frac{\alpha'}{2} \tau k_i lG(w,z_i)}
\,\partial_z \partial_wG(z,w)\, \partial_{\bar{z}}\partial_{\bar{w}}G(z,w)\nonumber\\
&&\label{3.2}
\end{eqnarray} 
where we have used the identity $\partial_z\partial_wG(z,w)=-\partial_wG(z,w)\partial_zG(z,w)$.

One integration, for example the one in $w$, can be easily performed by  using  the identity:
\begin{eqnarray}
&&e^{\frac{\alpha'}{2} \tau^2 q l G(w,z)}
\partial_z\partial_wG(z,w)\,\partial_{\bar{z}}\partial_{\bar{w}}G(z,w)=\frac{1}{\left(\frac{\alpha'}{2} \tau^2 ql -1\right)^2}\frac{\partial^2}{\partial z\partial\bar{z}}
|z-w|^{\alpha' \tau^2 ql -2}\label{3.3}
\end{eqnarray}

The integral in $w$ in \eqref{3.2}, with the use of \eqref{3.3} and after some  integration by parts,  can be easily evaluated in the soft region giving:
\begin{eqnarray}
&&I^{1100}_{1100}=
\int d^2 z \,\frac{\prod_{i=1}^ne^{\frac{\alpha'}{2} \tau q k_i G(z,z_i)}}{(\frac{\alpha'}{2} \tau^2 ql)^2\,(\frac{\alpha'}{2} \tau^2 ql -1)^2}
\frac{\partial}{\partial z}\frac{\partial}{\partial \bar{z}}\left[(2\pi)\alpha'\tau^2ql \prod_{i=1}^n e^{\tau lk_iG(z,z_i)}\right]+\Ord(\tau^0)~.\nonumber\\
&&\label{3.5}
\end{eqnarray}
where we have neglected all the boundary integrals. This integral, up to boundary terms, is equal to:
\begin{eqnarray}
&&I^{1100}_{1100}=-\frac{2\pi}{\frac{\alpha'}{2} \tau^2 ql \,(\frac{\alpha'}{2} \tau^2 ql -1)^2}\int d^2 z \,\partial_z\prod_{i=1}^ne^{\frac{\alpha'}{2} \tau q k_i G(z,z_i)}\,\partial_{\bar{z}} \prod_{i=1}^ne^{\frac{\alpha'}{2} \tau lk_i G(z,z_i)}\nonumber\\
&&=-2\pi\left(\frac{\alpha'}{2}\right) \sum_{i,j=1}^n \frac{ (q k_i) \,(l k_j)}{ql}\int d^2 z \, \partial_zG(z,z_i)\,\partial_{\bar{z}}G_c(z,z_j)\prod_{s=1}^ne^{\frac{\alpha'}{2} \tau (q+l)k_s G(z,z_s)}+\Ord(\tau^0)\nonumber\\
&&
\end{eqnarray}
The integral for the terms in the sum having $i=j$ is easily evaluated giving:
\begin{eqnarray}
&&-2\pi\sum_{i=1}^n \frac{ (q k_i) \,(l k_i)}{\tau\,ql k_i(q+l) }
\int d^2 z \,\partial_z e^{\frac{\alpha'}{2} \epsilon (q +l) k_i G(z,z_i)}\, \partial_{\bar{z}} G(z,z_i)\,\Big[1+\sum_{s\neq i}\frac{\alpha'}{2}\tau\,k_s(l+q)G(z,z_s)+\Ord(\tau^2)\Big] \nonumber\\
&& = -(2\pi)^2 \sum_{i=1}^n  \frac{(q k_i) \,(l k_i)}{\tau\,ql\,k_i(q+l)}\Big[\Lambda^{\alpha' \tau(q +l) k_i}+ \int d^2 z \,e^{\frac{\alpha'}{2}\tau (q +l) k_i G(z,z_i)}\, \pi \delta(z-z_i)\Big]+\Ord(\tau^0)\nonumber\\
&&= -(2\pi)^2 \sum_{i=1}^n  \frac{(q k_i) \,(l k_i)}{\tau\,ql\,k_i(q+l)}+\Ord(\tau^0)
\end{eqnarray}    
where $\Lambda$ is a cut-off regularizing the integral for large value of $z$ and  $e^{\frac{\alpha'}{2}\epsilon (q +l) k_i G(z_i,z_i)}=0$. In getting the result we have used the identities $\partial_z\partial_{\bar{z}}G(z,w)=\pi \delta(z-w)$ and $\int d^2z \delta(z-w)=2$.  

The integral for the terms of the sum with $i\neq j$ is computed in the same way and one can see that it is subleading in $\tau$.
{The terms in Eq. \eqref{Mn+2} containing $I^{1100}_{0011}$ therefore turn out to be:}
\ea{
N_0^2\,\varepsilon_q^{\mu\rho}\,\varepsilon_l^{\nu\sigma}&\Big(\eta_{\mu\nu}-\frac{\alpha'}{2}\tau^2\, l_\mu q_\nu\Big)\Big(\eta_{\rho\sigma} -\frac{\alpha'}{2} \tau^2 \,l_\rho q_\sigma\Big) I^{1100}_{1100}
\nonumber \\
&=-(2\pi)^2N_0^2 \varepsilon_q^{\mu\nu}\varepsilon_{l\,\mu\nu}\,\sum_{i=1}^n  \frac{(q k_i) \,(l k_i)}{\tau\,ql\,k_i(q+l)}+\Ord(\tau^0)~.
}
The next integrals that we consider are those with three derivative of the Green's function $G_c(z,w)$. They are:
\ea{
\frac{\alpha'}{2} N_0^2 \,\varepsilon_q^{\mu\rho}\varepsilon_l^{\nu\sigma}\,\tau\, \sum_{i=1}^n \Big[\eta_{\mu\nu} l_\rho k_{i\sigma} I^{1100}_{1001_i}+\eta_{\mu\nu}q_\sigma k_{i\rho} I^{1100}_{011_i0}+\eta_{\rho\sigma}l_\mu k_{i\nu} I^{1001_i}_{1100} +\eta_{\rho\sigma}q_\nu k_{i\mu} I^{011_i0}_{1100}\Big]
}
By exchanging the integration variables $(z,w)$ with their complex conjugate we get the following identities:
\begin{eqnarray}
I^{1100}_{1001_i}=I^{1001_i}_{1100}~~;~~I^{1100}_{011_i0}= I^{011_i0}_{1100}
\end{eqnarray} 
Furthermore, by exchanging $l$ with $q$ and the integrated variables $z$ with $w$, we get the further relation:
\begin{eqnarray}
I^{1100}_{1001_i}\Big|_{l\leftrightarrow q}=I^{1100}_{011_i0}
\end{eqnarray}
We have, therefore, only one integral to evaluate whose leading contribution in the $\tau$ expansion turns out to be:
\begin{eqnarray}
I^{1100}_{1001_i}=\frac{2}{\alpha'}\,\frac{(2\pi)^2\,(k_iq)}{\tau^2\,ql\,k_i(l+q)}+\Ord(\tau^{-1})\Rightarrow I^{1100}_{011_i0}=\frac{2}{\alpha'}\,\frac{(2\pi)^2\,(k_il)}{\tau^2\,ql\,k_i(l+q)}+\Ord(\tau^{-1})
\end{eqnarray}

The integrals with a double pole for $z\simeq w$ are of two different typologies:
\begin{eqnarray}
-\frac{\alpha'}{2}\,N_0^2\, \varepsilon_q^{\mu\rho}\varepsilon_l^{\nu\sigma}\sum_{i,j=1}^n\Big[k_{i\rho}\,k_{j\sigma}\eta_{\mu\nu} I^{1100}_{001_i1_j}+k_{i\mu}k_{j\nu} \eta_{\rho\sigma} I_{1100}^{001_i1_j}\Big]
\end{eqnarray}
The integral $I^{1100}_{001_i1_j}$ can be computed with the same procedure followed for $I^{1100}_{1100}$ which consists to integrate by part several times to get the double derivative of the Green's function which   
is equal to the delta-function. The result of the calculation gives:
\begin{eqnarray}
I^{1100}_{001_i1_j}=-\frac{(2\pi)^2\,\delta_{ij}}{\frac{\alpha'}{2} \tau\,k_i(q+l)(\frac{\alpha'}{2} \tau ql -1)} \Lambda^{\alpha'\tau k_i(q+l)}+\Ord(\tau^0)= \frac{(2\pi)^2\,\delta_{ij}}{\frac{\alpha'}{2} \tau\,k_i(q+l)}+\Ord(\tau^0) 
\end{eqnarray}
$I_{1100}^{001_i1_j}$ is the complex conjugate of $I^{1100}_{001_i1_j}$  and being $z,$ and $w$ integrate  variables we get:
\begin{eqnarray}
I_{1100}^{001_i1_j}=I^{1100}_{001_i1_j}
\end{eqnarray}
The second typology of integrals with a double pole for $z=w$ is:
\begin{eqnarray}
&&\Big(\frac{\alpha'}{2}\Big)^2N_0^2\epsilon_q^{\mu\rho}\epsilon_l^{\nu\sigma}\,\tau^2\,\sum_{i,j=1}^n\Big[l_\mu k_{j\nu}\Big(l_\rho \,k_{i\sigma} I^{1001_j}_{1001_i} +q_\sigma k_{i\rho} I^{1001_j}_{011_i0}\Big) +q_\nu k_{j\mu}\Big( l_\rho k_{i\sigma} I^{011_j0}_{1001_i} +q_\sigma k_{i\rho}I^{011_j0}_{011_i0}\Big)\Big]\nonumber\\
&&
\end{eqnarray}
We first observe that the exchange of the integration variables $z$ and $w$ together with the substitutions 
 $l \leftrightarrow q$ determines the following identity:
\begin{eqnarray}
I^{1001_j}_{1001_i}\Big|_{l\leftrightarrow q}=I^{011_j0}_{011_i0}~~;~~I^{1001_j}_{011_i0}\Big|_{l\leftrightarrow q}=I^{011_j0}_{1001_i}
\end{eqnarray}
The explicit calculation of the two independent integrals gives:
\begin{eqnarray}
I^{1001_j}_{1001_i}=-I^{1001_j}_{011_i0}=\Big(\frac{2}{\alpha'}\Big)^2\frac{(2\pi)^2\,\delta_{ij}}{\tau^3\,ql\,k_i(q+l)}+\Ord(\tau^{-2})
\end{eqnarray}
The integrals with only one derivative of $G_c(z,w)$ are:
\begin{eqnarray}
\Big(\frac{\alpha'}{2}\Big)^2 N_0^2 \varepsilon_q^{\mu\rho}\varepsilon_l^{\nu\sigma}\,\tau\!\!\!\sum_{i,j ,m=1}^n
\Big[&&k_{i\rho}k_{j\sigma} \Big( q_\nu k_{m\mu}I^{011_m0}_{001_i1_j}+l_\mu k_{m\mu} I^{1001_m}_{001_i1_j}\Big)\nonumber\\
+&&k_{i\mu}k_{j\nu} \Big( q_\sigma k_{m\rho}I^{001_11_j}_{011_m0}
+l_\rho k_{m\sigma} I^{001_i1_j}_{1001_m}\Big)\Big]
\end{eqnarray}
We now observe that the replacements $(z,\,q,\,i)\leftrightarrow (w,\,l,\,j)$ determine:
\begin{eqnarray}
I^{011_m0}_{001_i1_j}\Big|^{i\leftrightarrow j}_{l\leftrightarrow q} = I^{1001_m}_{00i_i1_j}~~;~~I^{001_i1_j}_{011_m0}\Big|^{i\leftrightarrow j}_{l\leftrightarrow q} = I^{001_i1_j}_{1001_m}
\end{eqnarray}
Furthermore, exchanging the integration variables with their complex conjugate we get:
\begin{eqnarray}
I^{011_m0}_{001_i1_j}=I_{011_m0}^{001_i1_j}
\end{eqnarray}
and, therefore,  there is only one integral to compute. Explicit calculation gives:
\begin{eqnarray}
I^{011_m0}_{001_i1_j}=I_{011_m0}^{001_i1_j}=\Big(\frac{2}{\alpha'}\Big)^2\frac{(2\pi)^2 \,\delta_{ij}\,\delta_{im}}{\tau^2\,k_i(q+l)\,k_iq}+\Ord(\tau^{-1})
\end{eqnarray}
which also implies:
\begin{eqnarray}
 I^{1001_m}_{00i_i1_j} = I^{001_i1_j}_{1001_m}=\Big(\frac{2}{\alpha'}\Big)^2\frac{(2\pi)^2 \,\delta_{ij}\,\delta_{im}}{\tau^2\,k_i(q+l)\,k_il}+\Ord(\tau^{-1})
\end{eqnarray}
The last integral to compute is the one without a pole for $z\simeq w$:
\begin{eqnarray}
\Big(\frac{\alpha'}{2} \Big)^2 N_0^2\,\varepsilon_q^{\mu\rho}\varepsilon_l^{\nu\sigma}\sum_{i,j,m,p=1}^n k_{i\mu}\,k_{j\nu}\,k_{m\rho}\,k_{p\sigma} \,I^{001_i1_j}_{001_m1_p} 
\end{eqnarray}
For such an integral, the exponential factor $e^{\frac{\alpha'}{2} \tau^2 lq G_c(z,w)}$ can be expanded  because any pole in $\tau$ can arise from the region of the complex plane where the two Koba-Nielsen variables $z$ and $w$ are pinched.   The integral becomes:
\begin{eqnarray}
&&\sum_{i,j,m,p=1}^n k_{i\mu}\,k_{j\nu}\,k_{m\rho}\,k_{p\sigma}\,I^{001_i1_j}_{001_m1_p} =\sum_{i,j,m,p=1}^n k_{i\mu}\,k_{j\nu}\,k_{m\rho}\,k_{p\sigma}\,\int d^2 z d^2 w\Big[1+\frac{\alpha'}{2}\tau^2 ql G_c(z,w)\Big]\nonumber\\
&&\times\prod_{i=1}^ne^{\frac{\alpha'}{2}\tau k_iq G_c(z,z_i)}\prod_{i=1}^ne^{\frac{\alpha'}{2} \tau k_i lG_c(w,z_i)}\partial_zG_c(z,z_i)\,\partial_wG_c(z,z_j)\,\partial_{\bar{z}}G_c(z,z_m)\,\partial_{\bar{w}}G_c(z,z_p)+\Ord(\tau^4)\nonumber\\
&&\label{3.25}
\end{eqnarray}
The leading contribution is exactly the same as two consecutive soft limits in the momenta $q$ and $l$ . The two integrals are decoupled and they have been computed in Ref.\cite{DiVecchia:2015oba}. The result is here quoted:
\begin{eqnarray}
(2\pi)^2\Big(\frac{2}{\alpha'}\Big)^2\Bigg[\sum_{i,j=1}^n \frac{k_i^\mu\,k_i^\rho}{\tau k_iq}\,\frac{k_j^\nu\,k_j^\sigma}{\tau k_jl}
-i \sum_{j=1}^n \frac{k_j^\nu\,k_j^\sigma}{\tau k_jl} \,\sum_{i=1}^n\frac{q_\tau k_i^\mu\, J_i^{\rho\tau}}{k_jq} -i\sum_{i=1}^n \frac{k_i^\mu k_i^\rho }{\tau k_iq} \,\sum_{j=1}^n \frac{l_\tau k_j^\nu J_j^{\sigma\tau}}{k_jl}\Bigg]+\Ord(\tau^0)\nonumber\\
\end{eqnarray}
Here $J_i^{\rho\tau}$ is the orbital angular momentum.

The subleading term in the $\tau$-expansion of \eqref{3.25} can be evaluated by using the same procedure followed  with the other integrals. The main contribution comes out  from the terms of the sum having $i=j=m=p$ and it turns out to be:
\begin{eqnarray}
-(2\pi)^2 \Big(\frac{2}{\alpha'}\Big)^2 \sum_{i=1}^n\frac{ k_i^\mu\,k_i^\nu\,k_i^\rho\,k_i^\sigma\,(ql)}{\tau k_il\,k_iq\,k_i(l+q)} +\Ord(\tau^0)
\end{eqnarray}

\section{Check of gauge invariance}
\label{gaugeinvariance}

On-shell amplitudes with external gauge particles like  gravitons or Kalb-Ramond fields, are  gauge invariant also in the infrared region  where the massless gauge particles carry low momenta. Gauge invariance is, therefore, a consistency condition that the  soft amplitude  derived in Eq. \eqref{8.3} must satisfy. It requires that a such $n+2$-point amplitude is vanishing when the external polarizations are replaced by the corresponding momentum,   according to the relations:
\begin{eqnarray}
\epsilon_{q\nu}\,q_\mu\,\varepsilon_{l\rho\sigma} M^{\mu\nu;\rho\sigma}_{n+2}(\{k_i\}, \tau q,\,\tau l)=\epsilon_{q\mu}\,q_\nu\,\varepsilon_{l\rho\sigma} M^{\mu\nu;\rho\sigma}_{n+2}(\{k_i\}, \tau q,\,\tau l)=0\label{4.1}
\end{eqnarray}
with $q_\mu\epsilon_q^\mu=0$ and similar relations for the other massless particle carrying momentum $l$. In the following, as consistency check of the string calculation,  we will show that indeed \eqref{8.3} satisfies these relations. It will be  sufficient to prove the condition \eqref{4.1}  only for the particle carrying $q$ momentum, because the symmetry of the amplitude in the exchange of $q$ with $l$ guarantees  that such constraints are also satisfied by the other soft momentum.

Equation \eqref{4.1}  when imposed on $M_1$ as written in \eqref{4.2},  and $M_2$ and $M_3$ as given in \eqref{2.6} and \eqref{2.7},   gives:
\ea{
& \sum_{i=1}^n \frac{\epsilon_{q\nu}q_\mu}{ k_i (q+l)}\Big[\frac{M_1^{\mu\nu}}{ql}+\frac{M_2^{\mu\nu}}{(k_iq)(k_il)} + M_3^{\mu \nu}\Big]M_n
=\sum_{i=1}^n
\Big[ (\epsilon_q l) \frac{(k_i\epsilon_lk_i)}{k_il}+\frac{(q\epsilon_lq)}{ql}(k_i \epsilon_q)\Big]M_n
}
The second term is subleading in the $\tau$-expansion and should be neglected.

The same replacement $\varepsilon_q^{\mu \nu} \to \epsilon_{q\nu} q_\mu $ when imposed on the remaining terms of \eqref{8.3} gives:
\ea{
&
\Bigg[\sum_{j=1}^n
(\epsilon_q k_i) \sum_{i=1}^n\frac{k_i\varepsilon_lk_i}{ k_il}-i\sum_{i=1}^n \frac{k_i\varepsilon_lk_i}{ k_il} \sum_{j=1}^n \epsilon_{q\nu}q_\rho J_j^{\nu\rho}-i\sum_{i=1}^n  (\epsilon_qk_i)
\sum_{j=1}^n \varepsilon_{l\mu\nu}\frac{ l_\rho k_j^\mu J_j^{\nu\rho}}{k_jl}\Bigg]M_n\nonumber\\
&=-(\epsilon_q l)  \sum_{i=1}^n\frac{k_i\varepsilon_lk_i}{k_il}\,M_n+ \cdots
\label{giJ}
}
where we have used momentum conservation and conservation of angular momentum in the form $\sum_{i=1}^n J_i^{\nu\rho} M_n =0$, and the $\cdots$ denote terms of higher order in the $\tau$-expansion, which should be neglected.
The relevant terms in the two last expressions exactly cancels,
thus ensuring gauge invariance.
Similar considerations hold when we replace 
 $\varepsilon_{q\mu\nu}\rightarrow \epsilon_{q\mu}\,q_\nu$,
 making \eqref{8.3} fully gauge invariant.
 
 
\section{Field theory diagrammatica}
\label{diagrammatica}
In this section we explain how the different terms in our main result \eqref{8.3}
can be understood from a Feynman diagram perspective.
A diagrammatic analysis of the single-soft case can be found in~\cite{DiVecchia:2016amo}.
The important lesson to draw from the single-soft case is that
both direct emission diagrams (where the soft state is emitted directly from external lines), 
as well as indirect emission diagrams (where the soft state is emitted from internal lines)
contribute to the sub- and subsubleading soft behavior of massless closed states,
whereas the leading soft behavior is entirely given by the direct emission diagrams (Weinberg's theorem).

At the double-soft level, direct emission diagrams come in two classes, cf. Fig.~\ref{Fig1}; 
A) where the two soft states are emitted directly from two different external lines,
and B) where they are emitted directly from the same external line. 
We classify the (mixed) type of diagrams where one state is emitted directly and another is emitted indirectly, as indirect emission diagrams together with the ones where both are emitted indirectly, cf. Fig~\ref{FigIndEm}. Since all terms of our result \eqref{8.3} have a pole in $\tau = 0$,
meaning that at least one propagator goes on-shell in the doubles-soft limit $q, l \to 0$,
we can immediately conclude that no diagrams, where both states are emitted indirectly (Fig.~\ref{FigIndEm}, right panel) contribute to \eqref{8.3} (since such diagrams at the tree-level are finite in this limit).
We can, however, not immediately conclude from the pole structure, which terms
are of direct emission type or of indirect emission type, 
since the class B direct emission diagrams
contain both 
 four-point contact double-emission diagrams which contribute with a simple pole in the $q, l \to 0$, as well as 
diagrams with an additional pole inside the four-point `blob' of Fig.~\ref{Fig1}, right panel, that require a deeper analysis to understand, which we will here perform.

This analysis, besides being instructive, also provides a direct relation 
between our result~\eqref{8.3} and the effective field theory action of 
massless closed strings interacting with massive scalar particles.
(For an effective action of tachyons interacting with gravitons and dilatons,
see Eq.~(4.12) in Ref.~\cite{Tseytlin:2000mt} and the equation below it.)

\begin{figure}[t]
\begin{center}
\begin{tikzpicture}[scale=.30]
		\draw [thick]  (8.5,2) -- (9.4,5); 
		\draw [thick]  (10.4,0.8) -- (13.6,2.6); 
		\draw [thick]  (6.75, 0.8) -- (4.65,2.5);
		\draw [thick]  (4.65, 2.5) -- (1.95,4.6) node[near end, sloped,below] { \scriptsize $k_{i}$};
		\draw [thick]  (4.65, -2.5) -- (1.95,-4.6) node[near end, sloped, above] { \scriptsize $k_{j}$};
		\draw [thick]  (6.75,-0.8) -- (4.65,-2.5); 
		\draw [thick]  (10.4,-0.8) -- (13.7,-1.7); 
		\draw [thick]  (8.5,-2) -- (9.4,-5);
		\draw[snake=coil,segment length=4pt] (7.15,5) -- (4.65,2.5) node[midway,sloped,above] { \scriptsize $\varepsilon_q,\, q$};
		\filldraw [thick, fill=lightgray] (4.65,2.5) circle (0.5cm);
		\draw [snake=coil,segment length=4pt] (7.15, -5) -- (4.65,-2.5) node[midway,sloped,below] { \scriptsize $\varepsilon_l,\, l$};
		\filldraw [thick, fill=lightgray] (4.65,-2.5) circle (0.5cm);
		\draw (8.5,.1) node { \footnotesize $M_n$};
		\draw [thick] (8.5,0) circle (2cm);
		\begin{scope}[shift={(-3,0)}] 
		\filldraw [ thick] (13.0,3.9) circle (1pt);
		\filldraw [ thick] (13.9,3.3) circle (1pt);
		\filldraw [ thick] (14.6,2.6) circle (1pt);
		\filldraw [ thick] (13.0,-3.9) circle (1pt);
		\filldraw [ thick] (13.9,-3.3) circle (1pt);
		\filldraw [thick] (7,1) circle(1pt);
		\filldraw [thick] (6.8,0) circle(1pt); 
		\filldraw [thick] (7,-1) circle(1pt); 
		\draw (10,-7.5) node { \footnotesize Class A diagrams};
		\end{scope}
		\end{tikzpicture}
		\hspace{1.0in}	
		\begin{tikzpicture}[scale=.30]
		\draw [thick]  (8.5,2) -- (9.4,5); 
		\draw [thick]  (10.4,0.8) -- (13.6,2.6); 
		\draw [thick]  (10.4,-0.8) -- (13.7,-1.7); 
		\draw [thick]  (8.5,-2) -- (9.4,-5); 
		\draw [thick] (-3.5,0) -- (-0.5,0) node[midway, above]{ \scriptsize $k_i$};
		\draw [thick](1.5,0) -- (6.5,0)node[midway, below] {\scriptsize $k_i+l+q$}; 
		\draw [snake=coil,segment length=4pt](0.5,1) --(0.5,4) node[near end, left]{\scriptsize $\varepsilon_l,\,l$};
		\draw[snake=coil,segment length=4pt] (0.5,-1) -- (0.5,-4) node[near end, left] { \scriptsize $\varepsilon_q,\, q$};
		\filldraw [thick, fill=lightgray] (0.5,0) circle (1cm);
		\draw (8.5,.1) node { \footnotesize $M_n$};
		\draw [thick] (8.5,0) circle (2cm);
		\begin{scope}[shift={(-3,0)}] 
		\filldraw [ thick] (13.0,3.9) circle (1pt);
		\filldraw [ thick] (13.9,3.3) circle (1pt);
		\filldraw [ thick] (14.6,2.6) circle (1pt);
		\filldraw [ thick] (13.0,-3.9) circle (1pt);
		\filldraw [ thick] (13.9,-3.3) circle (1pt);
		\draw (7.5,-6.5) node { \footnotesize Class B diagrams};
		\end{scope}
		\end{tikzpicture}
\end{center}
\caption{The two classes of direct emission diagrams of two massless closed states with polarization and momenta $\varepsilon_q, q$ respectively $\varepsilon_l, l$ from external tachyon lines with momenta $k_i$, with $i=1, \ldots, n$ and $i\neq j$.
Diagrams where the intermediate exchanged particle is not a tachyon is considered part of the indirect emission diagrams, cf. Fig.~\ref{FigIndEm}.
$M_n$ denotes the $n$-point tachyon subamplitude with two (A) or one (B) external legs off shell.
}
\label{Fig1}
\end{figure}

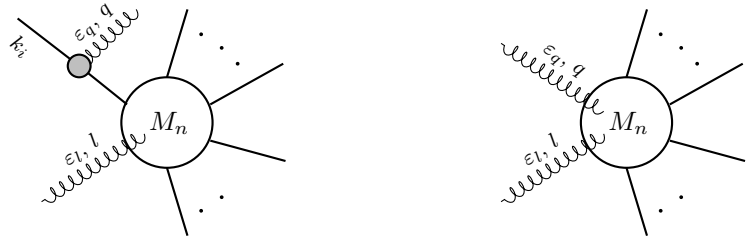
\begin{figure}[b]
\begin{center}
\begin{tikzpicture}[scale=.30]
		\draw [thick]  (8.5,2) -- (9.4,5); 
		\draw [thick]  (10.4,0.8) -- (13.6,2.6);
		\draw [thick]  (6.75, 0.8) -- (4.65,2.5); 
		\draw [thick]  (4.65, 2.5) -- (1.95,4.6) node[near end, sloped,below] { \scriptsize $k_{i}$};
		\draw [snake=coil,segment length=4pt]  
		(7.5,-0.5) -- (3,-3.5)
		node[midway, sloped, above] { \scriptsize $\varepsilon_l,\, l$};
		\draw [thick]  (10.4,-0.8) -- (13.7,-1.7); 
		\draw [thick]  (8.5,-2) -- (9.4,-5); 
		\draw[snake=coil,segment length=4pt] (7.15,5) -- (4.65,2.5) node[midway,sloped,above] { \scriptsize $\varepsilon_q,\, q$};
		\filldraw [thick, fill=lightgray] (4.65,2.5) circle (0.5cm);
		\draw (8.6,0) node { \footnotesize $M_{n}$};
		\draw [thick] (8.5,0) circle (2cm);
		\begin{scope}[shift={(-3,0)}] 
		\filldraw [ thick] (13.0,3.9) circle (1pt);
		\filldraw [ thick] (13.9,3.3) circle (1pt);
		\filldraw [ thick] (14.6,2.6) circle (1pt);
		\filldraw [ thick] (13.0,-3.9) circle (1pt);
		\filldraw [ thick] (13.9,-3.3) circle (1pt);
		\end{scope}
		\end{tikzpicture}
		\hspace{1.0in}	
		\begin{tikzpicture}[scale=.30]
		\draw [thick]  (8.5,2) -- (9.4,5); 
		\draw [thick]  (10.4,0.8) -- (13.6,2.6);
		\draw [snake=coil,segment length=4pt]  
		(7.5, 0.5) -- (3,3.5) 
		node[midway, sloped,above] { \scriptsize $\varepsilon_q,\, q$};
		\draw [snake=coil,segment length=4pt]  
		(7.5,-0.5) -- (3,-3.5)
		node[midway, sloped, above] { \scriptsize $\varepsilon_l,\, l$};
		\draw [thick]  (10.4,-0.8) -- (13.7,-1.7); 
		\draw [thick]  (8.5,-2) -- (9.4,-5);
		\draw (8.6,0) node { \footnotesize $M_{n}$};
		\draw [thick] (8.5,0) circle (2cm);
		\begin{scope}[shift={(-3,0)}] 
		\filldraw [ thick] (13.0,3.9) circle (1pt);
		\filldraw [ thick] (13.9,3.3) circle (1pt);
		\filldraw [ thick] (14.6,2.6) circle (1pt);
		\filldraw [ thick] (13.0,-3.9) circle (1pt);
		\filldraw [ thick] (13.9,-3.3) circle (1pt);
		\end{scope}
		\end{tikzpicture}
\end{center}
\caption{
The two types of indirect emission diagrams, where at least one of the massless closed string states is emitted from internal interactions of the $n$-point subamplitude.
Label descriptions as in Fig.~\ref{Fig1}.
}
\label{FigIndEm}
\end{figure}

\subsection{Class A direct emission contributions}
We begin by analyzing the almost trivial case of double-soft emission directly from two different external tachyons, labelled $i$ and $j$ with $i\neq j$, and $i,j = 1, \ldots, n$, cf. Class A of Fig.~\ref{Fig1} (left panel).
The intermediate (dashed) states carry momentum $k_i +q$ resp. $k_j + l$.
The propagators give the pole structure:
\ea{
\frac{1}{(k_i + q)^2 + m_i^2} \frac{1}{(k_j + l)^2 + m_j^2} 
= 
\frac{1}{(m_i^2 -m^2) + 2 (k_i q)} \frac{1}{ (m_j^2-m^2) + 2(k_j l)}
\label{double-pole}
}
where $-m^2 = k_i^2 = k_j^2$ is the tachyon mass, and we used that $q^2 = l^2 = 0$.
It is evident that if the intermediate state is a tachyon, 
the class A diagrams contribute with an overall double-pole 
$$\frac{1}{(k_i q) (k_j l)}$$ 
in the limit $q, l \to 0$.
If the intermediate states were both massless, or massless and tachyonic, 
the diagram is considered of indirect emission type, discussed later.
But we can immediately see from \eqref{double-pole} that if both 
intermediate states were massless, there would be no pole in the limit $q, l \to 0$, hence such diagrams do not contribute to \eqref{8.3}.

It is easy to calculate the full contribution of the class A diagrams at the tree level.
We notice that the three-point vertices (blobs in Fig.~\ref{Fig1}, left panel)
reduce to the three point polarization-stripped amplitude of 
two tachyons and one massless state, which is given by $M_3^{\mu \nu} = 2 \kappa_D \, k_i^\mu k_i^\nu$, assuming that the vertex is contracted with the polarization tensor $\varepsilon_q^{\mu \nu}$ and taken on shell, as is the case here. 
Hence the full contribution of the class A diagrams is given by:
\ea{
M_{n+2}^{(A)} = \kappa_D^2 \,  \varepsilon_{q, \mu \nu} \varepsilon_{l, \rho \sigma}
\sum_{i\neq j} \frac{k_i^\mu k_i^\nu}{(k_iq)} \frac{k_j^\rho k_j^\sigma}{(k_j l)}
M_n (k_1, \ldots, k_i + q, \ldots, k_j+l, \ldots, k_n) \, .
\label{MA}
}
In the limit $q, l \to 0$ we recognize
these contributions as the order $\tau^{-2}$-terms of \eqref{8.3} for $i \neq j$.
The corresponding $i=j$ contributions will naturally come from the class B direct
emission diagrams. 

The subleading contributions in the $q$ and $l$ expansion above 
are also represented in \eqref{8.3} and, less obviously, 
contained in the $J^{\mu \nu}$-terms. The other half of the $J^{\mu \nu}$-terms
come from indirect emission diagrams and ensure gauge invariance at the subleading order of the full amplitude.

\subsection{Class B direct emission contributions}

We start by noticing that the propagator shown in Fig.~\ref{Fig1}, right panel,
has a particular form in the soft limit, $q, l \to 0$:
\ea{
\frac{1}{(k_i + q + l)^2 + m_i^2 } =
\frac{1}{2 k_i (q+l) + 2 (ql) }
= \frac{1}{2 k_i(q+l)} - \frac{(ql)}{2 (k_i (q+l))^2} + \cdots
}
The pole-structure is that of the $M_1$ and $M_2$ terms of \eqref{8.3}.
We will furthermore show that the $(ql)$-term inside $M_2$, Eq.~\eqref{2.6},
is arising exactly due to the expansion above.

There are two ways to diagrammatically reproduce 
$M_1$ and $M_2$ terms of \eqref{8.3} from the class B diagrams:
We can either compute all diagrams contributing to the four-point `blob'
in Fig.~\ref{Fig1}, 
or more directly insert the four-point {string} amplitude for the `blob'
 by using that the intermediate off-shell leg is a scalar and 
 thus the four-point form-factor is a simple momentum-extension of the four-point amplitude.
 
 Since we here want to develop a diagrammatic understanding of all terms,
 we first proceed by the first approach and in the next section consider also the second approach 
 {that allows also to take in account the effect of the string corrections}.
 We start by listing the diagrams and effective point-interactions; i.e. Feynman vertices.
 
 \subsubsection{Three-point vertices}
 \label{3ptvertices}
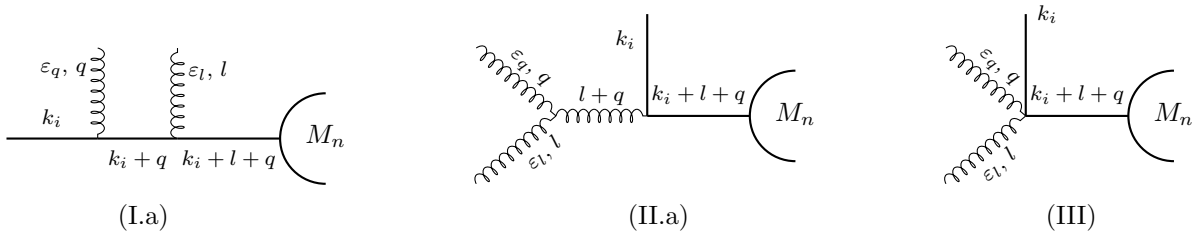
\begin{figure}[b]
	\begin{center}
	\begin{tikzpicture}[scale=.30]
		\draw[thick] (-5.5,0) -- (-1.5,0) node[midway, above]{ \scriptsize $k_i$};
		\draw[thick](2,0) -- (6.5,0)node[midway, below] {\scriptsize $k_i+l+q$} ;
		\draw[snake=coil,segment length=4pt](2,0) --(2,4) node[near end, right]{\scriptsize $\varepsilon_l,\,l$};
		\draw[snake=coil,segment length=4pt] (-1.5,4) -- (-1.5,0) node[near start, left] { \scriptsize $\varepsilon_q,\, q$};
		\draw[thick] (-1.5,0) -- (2,0) node[midway,below]{ \scriptsize $k_i+q$};
		\draw (8.5,.1) node { \footnotesize $M_n$};
		\draw [thick] (8.5,2)  arc[radius = 2cm, start angle= 90, end angle= 270];
		\draw (.5,-3.5) node { \footnotesize (I.a)};
		\end{tikzpicture}
		\hspace*{.5in}
		\begin{tikzpicture}[scale=.30]
		\draw[snake=coil,segment length=4pt] (-2,0) -- (2,0) node[midway, above]{ \scriptsize $l+q$};
		\draw[thick](2,0) -- (6.5,0)node[midway, above] {\scriptsize $k_i+l+q$} ;
		\draw[thick](2,0) --(2,4.5) node[near end, left]{\scriptsize $k_i$};
		\draw[snake=coil,segment length=4pt] (-5.5,3) -- (-2,0) node[midway, sloped,above] { \scriptsize $\hspace*{.01in}\varepsilon_q,\, q$};
		\draw [snake=coil,segment length=4pt] (-5.5,-3) -- (-2,0) node[midway, sloped,below] { \scriptsize $\hspace*{.2in}\varepsilon_l,\, l$};
		\draw (8.5,.1) node { \footnotesize $M_n$};
		\draw [thick] (8.5,2)  arc[radius = 2cm, start angle= 90, end angle= 270];
		\draw (2.5,-4.5) node { \footnotesize (II.a)};
		\end{tikzpicture}
		\hspace{0.5in}
		\begin{tikzpicture}[scale=.30]
		\draw[thick](2,0) -- (6.5,0)node[midway, above] {\scriptsize $k_i+l+q$} ;
		\draw[thick](2,0) --(2,4.5) node[right]{\scriptsize $k_i$};
		\draw[snake=coil,segment length=4pt] (-1.5,3) -- (2,0) node[midway,sloped,above] { \scriptsize $\varepsilon_q,\, q$};
		\draw [snake=coil,segment length=4pt] (-1.5,-3) -- (2,0) node[midway,sloped,below] { \scriptsize $\varepsilon_l,\, l$};
		\draw (8.5,.1) node { \footnotesize $M_n$};
		\draw [thick] (8.5,2)  arc[radius = 2cm, start angle= 90, end angle= 270];
		\draw (4,-4.5) node { \footnotesize (III)};
		\end{tikzpicture}	
\end{center}
\caption{The tree diagrams making up the class B diagrams of Fig.~\ref{Fig1}.}
\label{FigClassBDiagrams}
\end{figure}

 In Fig.~\ref{FigClassBDiagrams} we collect the tree diagrams that contribute
 to the class B diagrams of Fig.~\ref{Fig1}.

   There are three types of vertices in Fig.~\ref{Fig1}; 
  i) two-tachyon-one-massless, 
  ii) three-massless, 
  iii) two-tachyon-two-massless.

\begin{figure}[bt]
	\begin{center}
		\begin{tikzpicture}[scale=.30]
		\draw[thick] (-5.5,0) -- (-1.5,0) node[midway, above]{ \scriptsize $k_i$};
		\draw[thick](2,0) -- (6.5,0)node[midway, below] {\scriptsize $k_i+l+q$} ;
		\draw[snake=coil,segment length=4pt](2,0) --(2,4) node[near end, right]{\scriptsize $\varepsilon_l,\,l$};
		\draw[snake=coil,segment length=4pt] (-1.5,4) -- (-1.5,0) node[near start, left] { \scriptsize $\varepsilon_q,\, q$};
			\draw[snake=coil,segment length=4pt] (-1.5,0) -- (2,0) node[midway,below]{ \scriptsize $k_i+q$};
		\draw (8.5,.1) node { \footnotesize $M_n$};
		\draw [thick] (8.5,2)  arc[radius = 2cm, start angle= 90, end angle= 270];
		\draw (.5,-3.5) node { \footnotesize (I.b)};
		\end{tikzpicture}
		\hspace{0.5in}
		\begin{tikzpicture}[scale=.30]
		\draw[thick] (-2,0) -- (2,0) node[midway, above]{ \scriptsize $l+q$};
		\draw[thick](2,0) -- (6.5,0)node[midway, above] {\scriptsize $k_i+l+q$} ;
		\draw[thick](2,0) --(2,4.5) node[near end, left]{\scriptsize $k_i$};
		\draw[snake=coil,segment length=4pt] (-5.5,3) -- (-2,0) node[midway, sloped,above] { \scriptsize $\hspace*{.01in}\varepsilon_q,\, q$};
		\draw [snake=coil,segment length=4pt] (-5.5,-3) -- (-2,0) node[midway, sloped,below] { \scriptsize $\hspace*{.2in}\varepsilon_l,\, l$};
		\draw (8.5,.1) node { \footnotesize $M_n$};
		\draw [thick] (8.5,2)  arc[radius = 2cm, start angle= 90, end angle= 270];
		\draw (2.5,-4.5) node { \footnotesize (II.b)};
		\end{tikzpicture}
	\end{center}
\caption{Disallowed field theory tree diagrams of class B type.}
\label{Fig:WrongDiagrams}
\end{figure}
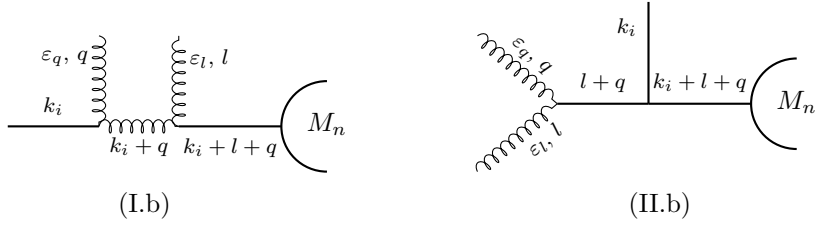

One could naively also write down the diagrams in Fig.~\ref{Fig:WrongDiagrams},
however, an inspection of the underlying three-point string
vertices shows that they have proper
field theory limit, see the end of App.~\ref{APPB}.
This is consistent with the effective field theory action in~\cite{Tseytlin:2000mt},
where the tachyon does not have three-point interactions with the graviton and dilaton involving only one tachyon. We therefore must disregard such diagrams 
and this turns out to be consistent, both with the double-soft theorem \eqref{8.3} and with the four-point amplitude of two tachyons and two massless states,
as shown in App.~\ref{App:4pt} and \ref{APPB}.

We derive the three-point vertices involved in Fig.~\ref{FigClassBDiagrams}
from the corresponding three-point string amplitudes, keeping particle symmetries manifest.

 Vertex (i) enters in diagrams (I.a) and (II.a). For diagram (I.a) it is sufficient to use the same form for the vertex as used for the class A diagrams, since the massless states are on shell. However, for diagram (II.a) we need an extension of vertex (i) taking into account the off-shellness of the massless state, and
 this is given by
 \begin{eqnarray}
V^{\alpha\beta}(p,k_1,k_2)= 2\kappa_D\frac{1}{2}(k_1-k_2)^{\alpha}\frac{1}{2} (k_1-k_2)^\beta
\label{10.4}
\end{eqnarray} 
where $k_1$, $k_2$ are the momenta of the tachyons and $p$ is the momentum of the massless state with polarization indices $\alpha \beta$.
Setting $k_1 =: k_i$ and $k_2 =: -k_1-p$ we get:
\begin{eqnarray}
V^{\alpha\beta}(p,k_i,\,-k_i-p)= 2\kappa_D(k_i+\frac{1}{2}p)^{\alpha} (k_i+\frac{1}{2}p)^\beta \, .
\label{Vonemassless}
\end{eqnarray} 
Eq.~\eqref{10.4} is simply the polarization-stripped 3-point string amplitude of two tachyons and one massless state, while \eqref{Vonemassless} is the promotion of it to a three-point vertex, where the $p$-terms only contribute if the massless state is internal (off shell), since $p_\alpha \varepsilon_p^{\alpha \beta} = 0$.

Vertex 
(ii)
for three-massless states is obtained in the same way and given by:
\ea{
V^{\mu\nu;\,\rho\sigma;\,\alpha\beta}(p_1;\,p_2;p_3) = &
\frac{\kappa_D}{2}
\left[
\eta^{\mu\rho}(p_1-p_2)^\alpha +\eta^{\rho\alpha}(p_2-p_3)^\mu+\eta^{\alpha \mu}(p_3-p_1)^\rho
+ \alpha' p_1^\alpha p_2^\mu p_3^\rho
\right]\nonumber\\
&\times\left[\eta^{\nu\sigma}(p_1-p_2)^\beta +\eta^{\sigma\beta}(p_2-p_3)^\nu+\eta^{\beta \nu}(p_3-p_1)^\sigma
+ \alpha' p_1^\beta p_2^\nu p_3^\sigma
\right]
\label{Vthreemassless}
}
Also the $\alpha'$-terms could have been written more symmetrically, e.g. $p_1^{\alpha} \to \frac{1}{2}(p_1 - p_2)^\alpha$, however, since these terms will be neglected in what follows, we kept the short form above.

Finally, the four-point vertex (iii) of two tachyons and two massless states
can be extracted from the four-point amplitude, which we return to later.
We will instead first predict its leading soft behavior from the double-soft theorem \eqref{8.3}.

\subsubsection{Diagram I.a}

As already discussed, the contributions from diagram I.a of Fig.~\ref{FigClassBDiagrams}
can be obtained similarly to the class B diagrams, 
remembering that we here have to sum over both the $t$ and $u$ channels (where the order of the two massless states are exchanged). We thus get:
\ea{
M_{n+2}^{\rm (I.a)} =
\kappa_D^2 \varepsilon_{q, \mu \nu} 
\varepsilon_{l, \rho \sigma} 
\sum_{i=1}^n \frac{ k_i^{\mu}k_i^{\nu}}{(k_iq)} 
\frac{ (k_i+q)^{\rho}(k_i+q)^{\sigma}}{k_i(q+l) + (ql)}
M_n(k_1 \ldots, k_i + q + l, \ldots k_n) + (q \leftrightarrow l)
} 
Introducing the $\tau$-parameter by $q \to \tau \, q $ and $l \to \tau \, l$ and expanding in $\tau$, one finds:
\ea{
M_{n+2}^{\rm (I.a)} =
\kappa_D^2 \varepsilon_{q, \mu \nu} 
\varepsilon_{l, \rho \sigma} 
\sum_{i=1}^n 
&\Bigg [\frac{1}{\tau^2} \frac{ k_i^{\mu}k_i^{\nu}}{(k_iq)} 
\frac{ k_i^{\rho}k_i^{\sigma}}{(k_il)}
\left ( 1 - \tau \frac{(ql)}{k_i(q+l)} + \tau (q+l)^\alpha \frac{\partial}{\partial k_i^\alpha}\right )
\nonumber\\
 &
+ \frac{1}{\tau} \frac{k_i^\mu k_i^\nu}{k_iq} \frac{ (k_i^\rho q^\sigma + k_i^\sigma q^\rho)}{k_i(q+l)}
+ \frac{1}{\tau} \frac{k_i^\rho k_i^\sigma}{k_il} \frac{ (k_i^\mu l^\nu + k_i^\nu l^\mu)}{k_i(q+l)}
\Bigg ] M_n(k_i) + \Ord(\tau^0)
\label{MIa}
}
The order $\tau^{-2}$ term together with the term with the $k_i$-derivative
give exactly the $i=j$ contributions corresponding to the sum in \eqref{MA}
related to \eqref{8.3} as discussed there.
The term with the contraction $(ql)$ and the terms in the second line
reproduce exactly the $M_2$-term in \eqref{8.3}, cf. \eqref{2.6}.

\subsubsection{Diagram II.a}
 
 The contributions from diagrams II.a are given by:
 \ea{
M_{n+2}^{\rm (II.a)}
=
\varepsilon_{q,\mu\nu} \varepsilon_{l,\rho\sigma} \sum_{i=1}^n 
& \frac{ V^{\mu \nu ;\rho \sigma; \alpha \beta} (q, l, -q - l )}{(q+l)^2 } 
 \frac{V_{\alpha \beta} 
 (q+l, k_i, -k_i-q-l )}{(k_i + q + l)^2 + m_i^2}
 \nonumber \\
 &
 \times
 M_n(k_1, \ldots, k_i + q + l, \ldots, k_n)
 }
 
 Parts of calculating the soft limit of this expression require some on-shell manipulations, 
and we thus give some detail.

 The first vertex above, after using on-shell tracelessness conditions, takes the form:
\ea{
V^{\mu\rho;\,\nu\sigma;\,\alpha\beta}(q,l,-q-l)=
&2\kappa_D
\left(\frac{1}{2}\eta^{\mu\nu}(q-l)^\alpha +\eta^{\nu\alpha}l^\mu-\eta^{\alpha \mu}q^\nu\right)
\nonumber\\
&\times
\left(\frac{1}{2}\eta^{\rho\sigma}(q-l)^\beta +\eta^{\sigma\beta}l^\rho-\eta^{\beta \rho}q^\sigma\right)
}
To evaluate $M_n^{\rm (II.a)}$ in the soft limit, we again introducing $\tau$ through $q\to \tau\, q$ and $l\to \tau\, l$
 and expand in $\tau$, getting:
\ea{
M_{n+2}^{\rm (II.a)}=
&\frac{\kappa_D^2}{\tau} 
\sum_{i=1}^n
\frac{1}{4(ql)(k_i(q+l))}
\Bigg[
(\varepsilon_l\varepsilon_q)k_i(q-l)\,k_i(q-l)
\nonumber\\
&
+2[(l\varepsilon_q^t \varepsilon_l k_i)+ (l \varepsilon_q \varepsilon_l^t k_i) ]k_i (q-l)
- 2 [ (k_i \varepsilon_q^t \varepsilon_l q) +  (k_i \varepsilon_q \varepsilon_l^t q)] k_i (q-l)
\nonumber\\
&
+4(l\varepsilon_ql)(k_i\varepsilon_lk_i){-4(l\varepsilon_qk_i)(k_i\varepsilon_lq)}
{- 4 (k_{i}\varepsilon_q l)(q\varepsilon_l k_i)}+4(k_i\varepsilon_q k_i)(q\varepsilon_l q)\Bigg]M_n(k_i)+\Ord(\tau^0)
\label{MIIa}
}
The terms with $k_i(q-l)$ can be rewritten to cancel the $k_i(q+l)$ denominator.
In particular, the first term is rewritten as follows:
\ea{
&\sum_{i=1}^n\frac{(\varepsilon_l\varepsilon_q)}{4(ql)(k_i(q+l))}
k_i(q-l)\,k_i(q-l)
\nonumber \\
&
= \sum_{i=1}^n\frac{(\varepsilon_l\varepsilon_q)}{4(ql)(k_i(q+l))}
k_i(q-l)\,k_i[(q+l)-2l]
\nonumber\\
&
= \sum_{i=1}^n\frac{(\varepsilon_l\varepsilon_q)}{4(ql)}k_i(q-l)
-\sum_{i=1}^n\frac{(\varepsilon_l\varepsilon_q)k_i(2q-(q+l))(k_il)}{2(ql)(k_i(q+l))}
\nonumber \\
&
=-\sum_i\frac{(\varepsilon_l\varepsilon_q)(k_iq)(k_il)}{ (ql)k_i(q+l)}
- \frac{(\varepsilon_l\varepsilon_q)}{2}
}
where in the last step momentum conservation was invoked to get (on shell)  
{$\sum_i k_i (q-l)=-(q+l)(q-l) = 0$} and \mbox{$\sum_i k_il = - (ql)$}.
The latter term is subleading in the $\tau$-expansion, and will thus be neglected
in \eqref{MIIa}.

Next we reduce the following terms in a similar manner:
\ea{
\sum_{i=1}^n
\frac{2[(l\varepsilon_q^t \varepsilon_l k_i)+ (l \varepsilon_q \varepsilon_l^t k_i) ]k_i (q-l)}{4 (ql) (k_i (q+l))}
&= 
\sum_{i=1}^n\frac{ [(l\varepsilon_q^t \varepsilon_l k_i)+ (l \varepsilon_q \varepsilon_l^t k_i) ](k_iq)}{(ql)(k_i(q+l))}
+ \frac{  (l \varepsilon_q^t \varepsilon_l q)}{ (ql) }
\\
-\sum_{i=1}^n
\frac{2 [ (k_i \varepsilon_q^t \varepsilon_l q) +  (k_i \varepsilon_q \varepsilon_l^t q)]k_i (q-l)}{4 (ql) (k_i (q+l))}
&= 
\sum_{i=1}^n\frac{  [ (k_i \varepsilon_q^t \varepsilon_l q) +  (k_i \varepsilon_q \varepsilon_l^t q)](k_il)}{(ql)(k_i(q+l))}
+ \frac{  (l \varepsilon_q^t \varepsilon_l q)}{(ql) }
}
The latter terms of each equation are subleading in the $\tau$-expansion and 
will thus be neglected in \eqref{MIIa}.

Finally, using that $\varepsilon_{q/l}^{\mu\nu}$ are either both symmetric or antisymmetric
we have the identity:
\ea{
(l\varepsilon_qk_i)(k_i\varepsilon_lq)
+(k_{i}\varepsilon_q l)(q\varepsilon_l k_i)
= 2 (l\varepsilon_q k_i)(k_i\varepsilon_l q)
}

In total $M_{n+2}^{\rm (II.a)}$ reduces to
\ea{
M_{n+2}^{\rm (II.a)}=
\frac{\kappa_D^2}{\tau} 
&\sum_{i=1}^n
\frac{1}{(ql)(k_i(q+l))}
\Bigg[
-(\varepsilon_l\varepsilon_q)(k_iq)(k_il)
-2(l\varepsilon_qk_i)(k_i\varepsilon_lq)
\nonumber\\
&
+[(l\varepsilon_q^t \varepsilon_l k_i)+ (l \varepsilon_q \varepsilon_l^t k_i) ](k_iq)
+ [ (k_i \varepsilon_q^t \varepsilon_l q) +  (k_i \varepsilon_q \varepsilon_l^t q)] (k_i l)
\nonumber \\
&
+(k_i\varepsilon_q k_i)(q\varepsilon_l q)
+(l\varepsilon_ql)(k_i\varepsilon_lk_i)
\Bigg]M_n(k_i)+\Ord(\tau^0)
\label{MIIareduced}
}
This expression matches exactly the contribution of the $M_1$ term in \eqref{8.3}.

\subsubsection{Diagram III}

This diagram is distinguished from the other one by only 
having one intermediate propagator taking the form on shell $1/k_i(q+l)$,
and involving a four-point, rather than three-point, contact interaction.
This interaction vertex can be extracted from the four-point string amplitude
by careful considerations, which we provide in Appendix~\ref{APPB}.

Here instead we want to make the point that our 
diagrammatic considerations so far together with our main result \eqref{8.3}
provides a prediction or derivation 
of the four-point contact interaction vertex to leading order in the soft expansion; 
i.e. the only term in \eqref{8.3}
with pole structure $1/k_i(q+l)$ and unmatched by the preceding diagrams
(there are no other diagrams giving such a pole structure) is 
the $M_3$-term, thus:
\ea{
M_{n+2}^{\rm (III)}
= - \kappa_D^2 \varepsilon_{q,\mu \nu} \varepsilon_{l,\rho \sigma}
\sum_{i=1}^n \frac{k_i^\mu k_i^\rho \eta^{\nu \sigma} + k_i^\nu k_i^\sigma \eta^{\mu \rho}}{\tau \, k_i(q+l)} M_n(k_i)
+ \Ord(\tau^0)
}
The prediction for the vertex is thus:
\ea{
V_4^{\mu \nu;\rho \sigma} (\tau q, \tau l, k_i, - k_i - q - l) = - 2 \kappa_D (k_i^\mu k_i^\rho \eta^{\nu \sigma} + k_i^\nu k_i^\sigma \eta^{\mu \rho}) + \Ord(\tau)
}
This matches exactly the complete vertex derived from the four-point amplitude in App.~\ref{C.10} to leading order in $\tau$.

\subsection{Indirect emission diagrams}

As we have already remarked, only the indirect emission diagrams in the left panel of Fig.~\ref{FigIndEm} may be contributing to the double-soft theorem in \eqref{8.3} (the right panel diagrams might contribute to a possible higher order soft theorem).

The contribution from these diagrams can now be fixed by gauge invariance.
We write them in the form:
\ea{
M_{n+2}^{\rm Ind.} = 
 &\sum_{i=1}^n \frac{\varepsilon_{q,\mu \nu} V^{\mu \nu} (q, k_i, -k_i - q)}{2 k_i q}
\varepsilon_{l,\rho \sigma} N^{\rho \sigma} (l, k_i+q)
\nn
&
+
 \sum_{i=1}^n \frac{\varepsilon_{l,\mu \nu} V^{\mu \nu} (l, k_i, -k_i - l)}{2 k_i l}
\varepsilon_{q,\rho \sigma} N^{\rho \sigma} (q, k_i+l)
\label{MInd}
}
where $N$ is an $n+1$-point form factor, which is local in its first argument, i.e.
\ea{
N^{\rho \sigma} (q,k_i) = N^{\rho \sigma} (0,k_i) + q \cdot \partial_q N^{\rho \sigma} (0,k_i)  + \cdots
}
and we recall that there are no such contributions if the two massless states are antisymmetric B-fields. Thus $N^{\rho \sigma}$ is a symmetric tensor. 
This is consistent with the fact that the sum of direct emission diagrams to the subleading soft order are gauge invariant by them selves.

The numerators in \eqref{MInd} are basically the three-point amplitude of two tachyons and one massless state, given by $2 \kappa_D (k_i \varepsilon_{q/l} k_i)$. 
Therefore to subleading order in $\tau$ (shifting $q\to \tau q$ and $l\to \tau l$) we simply have:
\ea{
M_{n+2}^{\rm Ind.} = 
 & \frac{\kappa_D^2}{\tau}\sum_{i=1}^n \left [ \frac{(k_i \varepsilon_{q} k_i)}{k_i q}
\varepsilon_{l,\rho \sigma} 
+
 \frac{(k_i \varepsilon_{l} k_i)}{ k_i l}
\varepsilon_{q,\rho \sigma} \right ] N^{\rho, \sigma} (0, k_i)
+ \Ord(\tau^0)
}
Since the full amplitude to this order in $\tau$ has to be gauge invariant,
and since all terms computed from the direct emission diagrams, except for the $(q+l)^\alpha\partial/\partial k_i^\alpha$-terms in \eqref{MA} and \eqref{MIa}, were already considered in Sec.~\eqref{gaugeinvariance}, we can directly read off from there, that $N^{\rho\sigma}$ is fixed by gauge invariance of the amplitude be:
\ea{
N^{\rho \sigma} (0, k_i) = - \sum_{j=1}^n 
 k_j^\rho \frac{\partial}{\partial k_{j \sigma} }M_n(k_i)
}
These contribution, together with the $(q+l)^\alpha\partial/\partial k_i^\alpha$-terms from \eqref{MA} and \eqref{MIa} then add to form the $J^{\rho \sigma}$-terms appearing in the double-soft theorem \eqref{8.3}, exactly so to restore gauge invariance of the full amplitude (cf. Sec.~\ref{gaugeinvariance}).

\section{Soft theorem from on-shell factorization}
\label{Factorization}

It might have been noticed that the diagrammatic calculation in the
previous section was essentially a computation of the
four-point amplitude, taking the soft limit, diagram-by-diagram, 
of the two massless states.

In fact, looking back at Fig.~\ref{Fig1}, right panel, 
one observes that the soft limit $q,l \to 0$ puts
the intermediate tachyon propagator on shell.
The factorization theorem then tells that
the amplitude factorizes into the corresponding lower-point amplitudes,
one of them being the four-point amplitude of two tachyons and two massless states.
Thus, when one is only interested in the leading and subleading double soft behaviour,
it suffices to use the factorization theorem to extract
the residues of the singularities from propagators going on shell in the soft limit by using the lower-point amplitudes.
Specifically,
all class B diagrams can be summarized as:
\ea{
M_{n+2}^{\rm class B}
&= \sum_{i=1}^n
M_4 (\tau q, \tau l, k_i, - (k_i + \tau q + \tau l) )
\frac{1}{\tau (2k_i (q+l))} M_n (k_i + \tau q +  \tau l) + \Ord (\tau^0)
\label{MClassB}
}
From the double-soft theorem \eqref{8.3} we expect at the same time
that the four-point string amplitude above reduce in the soft limit to the
four-point field theory amplitude, since no $\alpha'$-term enter in \eqref{8.3}.

To make an independent and alternative 
check that these relations are indeed correct,
we can simply compute the full four-point string
amplitude explicitly and insert it in the above expression.
The calculation of the four-point string amplitude of two tachyons and two massless
states is given in the App.~\ref{App:4pt}.
As shown in the Appendix, the result can be brought into the exact form:
\ea{
	M_{4}(\tau q, \tau l, k_1, k_2) =&  -2 \kappa_D^2 \, \varepsilon_{q}^{\mu \nu} \varepsilon_{l}^{\rho \sigma}
\frac{\Gamma(\frac{u}{2})\Gamma(\frac{2+s}{2})\Gamma(\frac{t}{2})}{\Gamma(\frac{4-t}{2})\Gamma(\frac{2-s}{2})\Gamma(\frac{4-u}{2})} 
\frac{ (k_1 q) (k_2 q)}{ql}
	\nn
	&\times \Bigg[
	A_{\mu \rho} +  \frac{\alpha' \tau^2 }{2-s} 
	\Big ( (ql) \eta_{\mu \rho} - (q_{\mu} + l_{\mu})(q_{\rho} + l_{\rho}) \Big )
 \Bigg]
	\nn
	&\times \Bigg[
	A_{\nu \sigma} +  \frac{\alpha' \tau^2 }{2-s} 
	\Big ( (ql) \eta_{\nu \sigma} - (q_{\nu} + l_{\nu})(q_{\sigma} + l_{\sigma}) \Big )
 \Bigg] \, ,
 }
 with
\ea{
A_{\mu \rho} = \eta_{\mu \rho}  + \frac{k_{1\mu} k_{2\rho}}{k_1 q} 
 +
  \frac{k_{2\mu} k_{1\rho}}{ k_2 q} \, ,
}
and where
\ea{
s= \alpha' \tau^2 (ql) \, , \quad t =2+ \alpha' \tau  (qk_2) \, , \quad u =2+\alpha' \tau (k_1q) \, .
}

Remarkably, each factor of $\alpha'$ comes with at least one power $\tau$ in the above expression for the four-point amplitude. 
This shows that the double-soft limit of this amplitude effectively takes the expression to its field theory limit, specifically:
\ea{
M_4(\tau q, \tau l, k_1, k_2) 
&=
-2 \kappa_D^2 \, 
\frac{ (k_1 q) (k_2 q)}{ql}
\varepsilon_{q}^{\mu \nu} \varepsilon_{l}^{\rho \sigma} \, 
A_{\mu \rho} A_{\nu \sigma} 
+ \Ord (\tau^2) 
\label{M4taulimit}
\\
M_4(\tau q, \tau l, k_1, k_2)  &= 
-2 \kappa_D^2 \, 
\frac{ (k_1 q) (k_2 q)}{ql}
\varepsilon_{q}^{\mu \nu} \varepsilon_{l}^{\rho \sigma} \, 
A_{\mu \rho} A_{\nu \sigma} 
+ \Ord (\alpha') 
\, ,
 \label{FTA4pt}
}
both of which is the field theory expression for the two-tachyon-two-massless-closed-states amplitude.
(The order $\tau$-term also vanish due to momentum conservation.)

This feature, together with the fact that the single-soft theorem does not contain $\alpha'$-terms, explains why we do not find any $\alpha'$-terms in the double-soft theorem.

Finally, considering \eqref{MClassB} and using \eqref{M4taulimit}, we 
can after straightforward but tedious manipulations show that
the contribution from the class B diagrams to the double-soft limit 
can be written as:
\ea{
M_{n+2}^{\rm Class B}
= \kappa_D^2 \sum_{i=1}^n
\Bigg[
&\frac{1}{\tau^2} \frac{(k_i\varepsilon_q k_i)(k_i\varepsilon_l k_i)}{(k_iq)(k_iq)}
\nn
&
 +
\frac{1}{\tau} 
\left (
\frac{M_1}{ql}+\frac{M_2}{(k_i q)(k_il)} + M_3\right)\frac{1}{k_i(q+l)}
\Bigg]\,M_n(k_i)
+\Ord(\tau^0)
}
where $M_{1,2,3}$ are exactly as given in \eqref{2.5}-\eqref{2.7}.

In the same way, all class A diagrams can be, almost trivially, 
understood from on-shell factorization on to the three-point subamplitudes; an analysis that we will not make explicit here.

This thus provides us with a complete understanding of all terms in \eqref{8.3};
it consists of the two consecutive single-soft terms (containing the expected $\Ord(\tau^{-2})$ Weinberg terms on different legs), while the rest is fully given by the factorization on the four-point subamplitude as given above.

\section{Conclusion and discussion}

In this work we have derived directly from 
the string theory integral representation of the amplitude of two massless closed strings interacting with $n$-point closed strings tachyons, 
the soft behavior of the two massless states when simultaneously becoming soft; i.e. when their momentum goes to zero.

The result for two gravitons has already been established in~\cite{Chakrabarti:2017ltl},
and confirmed here independently in a slightly different, but on-shell equivalent form,
while all the other cases; i.e. of two dilatons, a graviton and a dilaton, and two Kalb-Ramond states, are here derived for the first time.
Furthermore our main expression \eqref{8.3} provides a unified description of all the soft theorems just mentioned. Having no explicit $\alpha'$ contribution, the generic result can thus be seen as the double-soft theorem for the double-copied Yang-Mills theory interacting with massive scalar particles. We expect that our result is readily generalizable to the interaction with any other hard states by including in the angular momenta operator, appearing in the subleading soft operator, also the spin angular momenta part.

One of the uses of soft theorems is to derive from them
the low-energy effective field theory action of the underlying theory.
In this spirit, we have provided also a diagrammatic analysis of the
terms appearing in the double-soft theorem \eqref{8.3}.
This establishes for the low-energy theory two things:
1) There are no three-point field theory interactions of two massless states interacting with one tachyon, and the effective field theory should be void from such (this agrees with Ref.~\cite{Tseytlin:2000mt}), 
and 2) there is a four-point contact interaction of two tachyons and two massless states, which we have derived both from the double-soft theorem, and from an analysis of the four-point amplitude.

Due to the high level of complexity in the analysis presented in this paper, 
we have here focused our attention on the leading and subleading double-soft behavior.
It would be very interesting to analyze also the next, i.e. subsubleading
$\tau^0$ order, where we expect much more complexity to appear, since given that 
the single-soft cases all have a subsubleading factorization behavior, 
it might turn out that the high level of complexity will organize compactly into a subsubleading double-soft theorem due to some putative unknown (hidden) symmetry. 
Having now a good handle on the first two orders, we hope that such an analysis will appear in a future work.

\subsection*{Acknowledgments} 

We thank Paolo Di Vecchia for initial collaboration and useful discussions on this work.
R.M. thanks  Mritunjay Verma for interesting discussions on the subject.

\appendix

\section{Four-point bosonic string amplitude of two closed tachyons and two closed massless string states}
\label{App:4pt}

The 2-tachyon-2-massless closed string amplitude is given at the tree-level by:
\begin{align}
M_{ttdd} = &C_0 N_0^4 \varepsilon_3^{\mu \nu} \varepsilon_4^{\rho \sigma} \int \frac{\prod_{i=1}^4 d^2 z_i}{dV_{abc}} \prod_{i<j} |z_i - z_j |^{\alpha' k_i k_j }
\nonumber \\
&
\times \Bigg [- \frac{\eta_{\mu \rho}}{(z_3-z_4)^2}
+ \frac{\alpha'}{2} \sum_{i,j=1}^2 \frac{k_{i\mu}k_{j\rho}}{(z_3-z_i)(z_4-z_j)}
+ \frac{\alpha'}{2} \sum_{i=1}^2 \frac{k_{4\mu}k_{i\rho}}{(z_3-z_4)(z_4-z_i)}
\nonumber \\
&
+ \frac{\alpha'}{2} \sum_{i=1}^2 \frac{k_{i\mu}k_{3\rho}}{(z_3-z_i)(z_4-z_3)}
- \frac{\alpha'}{2} \frac{k_{4\mu}k_{3\rho}}{(z_3-z_4)^2}
\Bigg]
\times
\Bigg [ \text{c.c. with } (\mu \to \nu, \rho \to \sigma)
\Bigg]
\end{align}
Here $k_1,k_2$ are the tachyon momenta.
To perform the integral, we set $z_1 = 0, \ z_2 = z , \ z_3 =1 , \ z_4 = w \to \infty$, thus $d V_{134} = d^2 z_1 d^2 z_3 d^2 z_4 / | w |^2 | 1- w|^2 \to 
 d^2 z_1 d^2 z_3 d^2 z_4 / | w |^4 $, and get:
\begin{align}
M_{ttdd} = &C_0 N_0^4 \varepsilon_3^{\mu \nu} \varepsilon_4^{\rho \sigma} \int d^2 z  |  z |^{\alpha' k_1k_2 } |  w |^{\alpha' k_1k_4 } |  1- z |^{\alpha' k_2k_3 } |  z-w |^{\alpha' k_2k_4 }  |  1- w |^{\alpha' k_3k_4 }
\nonumber \\
&
\times \Bigg [- \eta_{\mu \rho} 
+ \frac{\alpha'}{2} \sum_{i,j=1}^2 \frac{w k_{i\mu}k_{j\rho}}{(1-z_i)(1-\frac{z_j}{w})}
- \frac{\alpha'}{2} \sum_{i=1}^2 k_{4\mu}k_{i\rho}
\nonumber \\
&
+ \frac{\alpha'}{2} \sum_{i=1}^2 \frac{w k_{i\mu}k_{3\rho}}{(1-z_i)(1-\frac{1}{w})}
- \frac{\alpha'}{2} k_{4\mu}k_{3\rho}
\Bigg]
\times
\Bigg [ \text{c.c. with } (\mu \to \nu, \rho \to \sigma)
\Bigg]
\end{align}
where in the bracket we already took the limit $w \to \infty$, where it was consistent.
Notice that
\begin{align}
&\sum_{i,j=1}^2 \frac{w k_{i\mu}k_{j\rho}}{(1-z_i)(1-\frac{z_j}{w})}
+ \sum_{i=1}^2 \frac{w k_{i\mu}k_{3\rho}}{(1-z_i)(1-\frac{1}{w})}
= \sum_{i=1}^2 \sum_{j=1}^3 \frac{w k_{i\mu}k_{j\rho}}{(1-z_i)(1-\frac{z_j}{w})}
\nonumber \\
&=  \sum_{i=1}^2 \sum_{j=1}^3 \frac{w k_{i\mu}k_{j\rho}}{(1-z_i)}\left (1 + \frac{z_j}{w} + \cdots \right )
= - \sum_{i=1}^2\frac{w k_{i\mu}k_{4\rho}}{(1-z_i)} + \sum_{i=1}^2 \sum_{j=1}^3\frac{z_j k_{i\mu}k_{j\rho}}{(1-z_i)}
\end{align}
and we see that the first term vanishes on shell (as it must) due to $\varepsilon_4^{\rho\sigma} k_{4 \rho} = 0$.
We can also drop the term in the second sum where $j=1$, since $z_1 = 0$.
Also the following two terms vanish on shell:
\begin{align}
\varepsilon_4^{\rho\sigma}\left (-\sum_{i=1}^2 k_{4\mu}k_{i\rho}
- k_{4\mu}k_{3\rho} \right ) = 0
\end{align}
Notice also that for $w \to \infty $.
\begin{align}
|  w |^{\alpha' k_1k_4 }  |  z-w |^{\alpha' k_2k_4 }  |  1- w |^{\alpha' k_3k_4 }
\to |w|^{\alpha' (k_1 + k_2 + k_3)k_4 } = |w|^0 = 1
\end{align}
since $k_4^2=0$ on shell.
 Then we are left with the following expression:
 \begin{align}
M_{ttdd} = &C_0 N_0^4 \varepsilon_3^{\mu \nu} \varepsilon_4^{\rho \sigma} \int d^2 z  |  z |^{\alpha' k_1k_2 }  |  1- z |^{\alpha' k_2k_3 } \nonumber \\
&\times
\left [\eta_{\mu \rho}  
+ \frac{\alpha'}{2} \sum_{i=1}^2 \sum_{j=2}^3\frac{z_j k_{i\mu}k_{j\rho}}{(1-z_i)} \right ]
\left [
\eta_{\nu \sigma}  + 
 \frac{\alpha'}{2} \sum_{i=1}^2 \sum_{j=2}^3\frac{\bar{z}_j k_{i\mu}k_{j\rho}}{(1-\bar{z}_i)} \right ]
 \, .
\end{align}
Expanding the sum in the bracket gives:
\begin{align}
&\Bigg [
\eta_{\mu \rho}  
+ \frac{\alpha'}{2}
\left (
 z k_{1\mu} k_{2\rho} + k_{1\mu} k_{3\rho}
+ \frac{1}{1-z} (z k_{2\mu} k_{2\rho} +k_{2\mu}  k_{3\rho})
\right)
\Bigg ] 
\nonumber \\
&=\Bigg [
\eta_{\mu \rho}  
+ \frac{\alpha'}{2}\frac{1}{(1-z)} 
\left (
 z (1-z) k_{1\mu} k_{2\rho} + (1-z) k_{1\mu} k_{3\rho}
+ z k_{2\mu} k_{2\rho} +k_{2\mu}  k_{3\rho}
\right)
\Bigg ] 
\nonumber \\
&=\Bigg [
\eta_{\mu \rho}  
+ \frac{\alpha'}{2}\frac{1}{(1-z)} 
\left (
- z^2  k_{1\mu} k_{2\rho} 
+z \underbrace{(k_{1\mu} k_{2\rho} - k_{1\mu} k_{3\rho}+k_{2\mu} k_{2\rho})}_
{-k_{4\mu} k_{2\rho} - k_{1\mu}k_{3\rho}}
+ \underbrace{k_{1\mu} k_{3\rho} +k_{2\mu}  k_{3\rho}}_{-k_{4\mu} k_{3\rho}}
\right)
\Bigg ] 
\end{align}
where in the underbraced expressions we used momentum conservation and on-shell conditions. 

Rewriting it further by using momentum conservation and on-shell condition, $k_{3\rho} \to - (k_1 + k_2)_\rho$ and $k_{4 \mu} \to  - (k_1 + k_2)_\mu $ leads to the following simpler form:
\begin{align}
\Bigg [
\eta_{\mu \rho} - 
 \frac{\alpha'}{2} \left (
 k_{1\mu} k_{1\rho} +  k_{2\mu} k_{2\rho} \right )
 -
 \frac{\alpha'}{2}  k_{1\mu} k_{2\rho} (1-z) 
 -
 \frac{\alpha'}{2}  \frac{k_{2\mu} k_{1\rho}}{ (1-z) } 
\Bigg ] 
\end{align}

We now note that all integrals are in the form:
\begin{align}
	I_{n, \bar{n}} = \int d^2 z \, 
	z^{A} \, (1-z)^{B+n}\, \bar{z}^{A}\, (1-\bar{z})^{B+ \bar{n}} \, .
\end{align}
It can be shown, using integration techniques of KLT, that all such integrals can be rewritten into the form
\begin{align}
	I_{n, \bar{n}} = 
2\pi\,	\frac{\Gamma(-1-A-B-n)}{\Gamma(-A)\Gamma(-B-n)} \frac{\Gamma(1+A)\Gamma(1+B+\bar{n})}{\Gamma(2 + A+ B+ \bar{n})}
\end{align}
We notice that the left and right brackets, parametrized by $n$ respectively $\bar{n}$ factorize, thus the factorized structure of kinematic coefficients will also preserved.
By using the property $z \Gamma(z) = \Gamma(1+z)$, we can extract an overall function outside of the
brackets of the form:
\begin{align}
	C(A',B):=
	\frac{\Gamma(1-A'-B)}{\Gamma(1-A')\Gamma(1-B)} \frac{\Gamma(1+A')\Gamma(1+B)}{\Gamma(1 + A'+ B)}
	=: C_L C_R
\end{align}
where we made a redefinition of $A$, i.e. $A' = A + 2 = \frac{\alpha'}{2}k_3k_4$, upon using momentum conservation, on-shell conditions and that $m^2 = - 4/\alpha'$.
This prefactor has the property that $\lim_{\alpha'\to 0} C = 1$.

We write down each case of $n$ separately.

For $n=0$ we need the following rewriting to factorize in $C_L$:
\begin{align}
	\frac{\Gamma(-1-A-B)}{\Gamma(-A)\Gamma(-B)}
	&= 
	\frac{\Gamma(1-A'-B)}{\Gamma(2-A')\Gamma(-B)}
	= \frac{(-B)}{(1-A')}  \, C_L
\end{align}

Similarly, for $n=-1$ we need:
\begin{align}
	\frac{\Gamma(-A-B)}{\Gamma(-A)\Gamma(1-B)}
	&= 
	\frac{\Gamma(2-A'-B)}{\Gamma(2-A')\Gamma(1-B)}
	= \frac{\left ( 1-A'-B\right )}{1-A'}
	C_L
\end{align}

And for $n=1$ we need:
\begin{align}
	\frac{\Gamma(-2-A-B)}{\Gamma(-A)\Gamma(-B-1)}
	&= 
	\frac{\Gamma(-A'-B)}{\Gamma(2-A')\Gamma(-B-1)}
	= \frac{(-B-1)(-B)}{(-A'-B)(1-A')}C_L
\end{align}

For $\bar{n} = 0$ we need the following rewriting to factorize into $C_R$.
\begin{align}
	\frac{\Gamma(1+A)\Gamma(1+B)}{\Gamma(2 + A+ B)}
	=
	\frac{\Gamma(-1+A')\Gamma(1+B)}{\Gamma( A'+ B)}
	= \frac{(A'+B)}{(-1+A')(A')} C_R
\end{align}

Similarly, for $\bar{n}=-1$ we need:
\begin{align}
	\frac{\Gamma(1+A)\Gamma(B)}{\Gamma(1 + A+ B)}
	=
	\frac{\Gamma(-1+A')\Gamma(B)}{\Gamma(-1 + A'+ B)}
	=\frac{(-1+A'+B)(A'+B)}{(-1+A')(A')(B)}C_R
\end{align}

And finally for $\bar{n}=1$ we have:
\begin{align}
	\frac{\Gamma(1+A)\Gamma(2+B)}{\Gamma(3 + A+ B)}
	=
	\frac{\Gamma(-1+A')\Gamma(2+B)}{\Gamma(1 + A'+ B)}
	=
	\frac{(1+B)}{(-1+A')(A')}C_R
\end{align}

We are now ready to assemble all pieces into the amplitude, getting:
\begin{align}
	M_{ttdd} =& C_0 N_0^4 \varepsilon_3^{\mu \nu} \varepsilon_4^{\rho \sigma}
	(2\pi)\left [\frac{C(A',B)}{(1-A')^2} \right ] \frac{B(A'+B)}{A'}
	\nn
	&\times \Bigg[
	\eta_{\mu \rho} - 
 \frac{\alpha'}{2} \left (
 k_{1\mu} k_{1\rho} +  k_{2\mu} k_{2\rho} \right )
 -
 \frac{\alpha'}{2}  k_{1\mu} k_{2\rho} \frac{1+B}{A'+B} 
 +
 \frac{\alpha'}{2}  k_{2\mu} k_{1\rho} \frac{1-A'-B}{B}
 \Bigg]
 \nn
 	&\times \Bigg[
	\eta_{\nu \sigma} - 
 \frac{\alpha'}{2} \left (
 k_{1\nu} k_{1\sigma} +  k_{2\nu} k_{2\sigma} \right )
 -
 \frac{\alpha'}{2}  k_{1\nu} k_{2\sigma} \frac{1+B}{A'+B} 
 +
 \frac{\alpha'}{2}  k_{2\nu} k_{1\sigma} \frac{1-A'-B}{B}
 \Bigg]
\end{align}
where
$$
A' = \frac{\alpha'}{2} k_3 k_4 \, , \quad B = \frac{\alpha'}{2} k_2 k_3
$$
or in terms of the Mandelstam variables:
$$
s = \frac{\alpha'}{2} (k_3+k_4)^2 
 \, , \quad
 t = \frac{\alpha'}{2} (k_2+k_3)^2 
  \, , \quad
 u = \frac{\alpha'}{2} (k_2+k_4)^2 
 \, , \quad s+t+u = 4
 $$
 $$
 \Rightarrow \quad
 A' = \frac{s}{2} \, , \quad B = \frac{t-2}{2} \, ,\quad
 A'+B = \frac{s+t-2}{2} = \frac{2-u}{2}
 $$
We can write the amplitude as
\begin{align}
	M_{ttdd} =& C_0 N_0^4 \varepsilon_3^{\mu \nu} \varepsilon_4^{\rho \sigma}
	(4\pi)\left [\frac{C(s,t,u)}{(2-s)^2} \right ] \frac{(t-2)(2-u)}{s}
	\nn
	&\times \Bigg[
	\eta_{\mu \rho} - 
 \frac{\alpha'}{2} \left (
 k_{1\mu} k_{1\rho} +  k_{2\mu} k_{2\rho} \right )
 +
 \frac{\alpha'}{2}  k_{1\mu} k_{2\rho} \, \frac{t}{u-2} 
 +
 \frac{\alpha'}{2}  k_{2\mu} k_{1\rho} \, \frac{u}{t-2}
 \Bigg]
 \nn
 	&\times \Bigg[
	\eta_{\nu \sigma} - 
 \frac{\alpha'}{2} \left (
 k_{1\nu} k_{1\sigma} +  k_{2\nu} k_{2\sigma} \right )
 +
 \frac{\alpha'}{2}  k_{1\nu} k_{2\sigma} \,\frac{t}{u-2}  
 +
 \frac{\alpha'}{2}  k_{2\nu} k_{1\sigma} \,\frac{u}{t-2}
 \Bigg]
\end{align}
where
\begin{align}
C(s,t,u):= C\left(\frac{s}{2},\frac{t-2}{2}\right)
= 
\frac{\Gamma(\frac{u}{2})\Gamma(\frac{2+s}{2})\Gamma(\frac{t}{2})}{\Gamma(\frac{4-t}{2})\Gamma(\frac{2-s}{2})\Gamma(\frac{4-u}{2})} \, ,
\end{align}
and we note that:
$$
t-2 = \alpha' k_2 k_3 \, , \quad u-2 =-\alpha'k_3(k_2+k_4)=\alpha' k_1 k_3
$$
$$
\Rightarrow \quad  \frac{(t-2)(2-u)}{s} =- \alpha'\frac{(k_2k_3)(k_1k_3)}{k_3k_4}
$$
So with
$$
C_0 = \frac{8\pi}{\alpha'}\left (\frac{2\pi}{\kappa_D}\right )^2 \, , \quad
N_0 = \frac{\kappa_D}{2\pi}
$$
we have
\begin{align}
	M_{ttdd} =&   \varepsilon_3^{\mu \nu} \varepsilon_4^{\rho \sigma}
(-2 \kappa_D^2 )\left [\frac{C(s,t,u)}{(1-\frac{s}{2})^2} \right ] \frac{(k_2k_3)(k_1k_3)}{k_3k_4}
	\nn
	&\times \Bigg[
	\eta_{\mu \rho} - 
 \frac{\alpha'}{2} \left (
 k_{1\mu} k_{1\rho} +  k_{2\mu} k_{2\rho} \right )
 +
 \frac{t}{2}  \frac{k_{1\mu} k_{2\rho}}{k_1k_3} 
 +
 \frac{u}{2}  \frac{k_{2\mu} k_{1\rho}}{k_2k_3}
 \Bigg]
 \nn
 	&\times \Bigg[
	\eta_{\nu \sigma} - 
 \frac{\alpha'}{2} \left (
 k_{1\nu} k_{1\sigma} +  k_{2\nu} k_{2\sigma} \right )
 +
 \frac{t}{2}  \frac{k_{1\nu} k_{2\sigma}}{k_1k_3} 
 +
 \frac{u}{2}  \frac{k_{2\nu} k_{1\sigma}}{k_2k_3}
 \Bigg]
\end{align}

We can immediately derive the field theory limit of this, since
In the limit $\alpha'\to 0$ we know that 
$s\to 0$, and $t,u \to 2$ and thus
$C(s,t,u)\to 1$%
\footnote{This follows e.g. from  
$\frac{\Gamma (1+z)}{\Gamma(1-z)} =- \frac{\Gamma(z)}{\Gamma(-z)} = 1-2z \gamma + \Ord(z^2)$, where $\gamma$ 
is the Euler-Mascheroni constant.}.
Thus the field theory result is:
\begin{align}
	\lim_{\alpha'\to 0} M_{ttdd} =&   
	-2 \kappa_D^2 \, 
\frac{(k_2k_3)(k_1k_3)}{k_3k_4}
\varepsilon_{3}^{\mu \nu} \varepsilon_{4}^{\rho \sigma} \, 
A_{\mu \rho} A_{\nu \sigma} \, ,
 \label{App:FTA4pt}
\end{align}
with
\ea{
A_{\mu \rho} = \eta_{\mu \rho}  + \frac{k_{1\mu} k_{2\rho}}{ k_1 k_3} 
 +
  \frac{k_{2\mu} k_{1\rho}}{ k_2 k_3} \, .
}
This agrees, up to an overall sign, 
with the field theory result given in Eq.~(2.2) of Ref.~\cite{KoemansCollado:2019ggb}.

We can, in fact, separate the field theory part from the string corrections as follows.
We 
first use momentum conservation to rewrite $t = 2 - \alpha' (k_1k_3 + k_3k_4)$ and $u= 2-\alpha'(k_2k_3 + k_3k_4)$, and then we can algebraically rewrite
the amplitude into the following form (using in particular $\frac{1}{1-\frac{s}{2}} = 1 + \frac{s}{2-s}$): 
\ea{
	M_{ttdd}  =& -2 \kappa_D^2 \, \varepsilon_{3}^{\mu \nu} \varepsilon_{4}^{\rho \sigma}
\frac{\Gamma(\frac{u}{2})\Gamma(\frac{2+s}{2})\Gamma(\frac{t}{2})}{\Gamma(\frac{4-t}{2})\Gamma(\frac{2-s}{2})\Gamma(\frac{4-u}{2})} 
 \frac{(k_2k_3)(k_1k_3)}{k_3 k_4}
	\nn
	&\times \Bigg[
	A_{\mu \rho} +  \frac{\alpha'}{2-s} 
	\Big ( (k_3k_4) \eta_{\mu \rho} - (k_{1\mu} + k_{2\mu})(k_{1\rho} + k_{2\rho}) \Big )
 \Bigg]
	\nn
	&\times \Bigg[
	A_{\nu \sigma} +  \frac{\alpha'}{2-s} 
	\Big ( (k_3k_4) \eta_{\nu \sigma} - (k_{1\nu} + k_{2\nu})(k_{1\sigma} + k_{2\sigma}) \Big )
 \Bigg]
 }
Using momentum conservation we can further rewrite $(k_{1\mu} + k_{2\mu}) = - (k_{3\mu} + k_{4\mu})$, getting the form used in the main text, where the notation $k_3 = q$, $k_4 = l$ and $\varepsilon_3 = \varepsilon_q$, $\varepsilon_4 = \varepsilon_l$ is used.

\section{ Four-point amplitude from string inspired Feynman diagrams}
\label{APPB}

In this appendix we show how to reproduce the field theory result of the four-point 
amplitude of two tachyons and two massless closed strings, calculated in the previous appendix, through a diagrammatic field theory like calculation, 
thereby establishing the Feynman rules and vertices for the field theory limit of the theory. Our aim here is in particular to fix the four-point vertex (on-shell contact term) of two tachyons and two massless states.

\begin{figure}[b]
	\begin{center}
	\begin{tikzpicture}[scale=.30]
		\draw[thick] (-2,0) -- (2,0) node[midway, above]{ \scriptsize $k_1+q$};
		\draw[snake=coil,segment length=4pt] (-5.5,3) -- (-2,0) node[midway, sloped,above] { \scriptsize $\hspace*{.01in}\varepsilon_q,\, q$};
		\draw[snake=coil,segment length=4pt] (5.5,3) -- (2,0) node[midway, sloped,above] { \scriptsize $\hspace*{.01in}\varepsilon_l,\, l$};
		\draw[thick] (-5.5,-3) -- (-2,0) node[midway,below] {\scriptsize $ k_1$};
		\draw[thick] (5.5,-3) -- (2,0) node[midway,below] {\scriptsize $ k_2$};
		\draw (0.7,-4.5) node { \footnotesize (a)};
		\end{tikzpicture}
		\hspace*{.5in}
		\begin{tikzpicture}[scale=.30]
		\draw[snake=coil,segment length=4pt] (-2,0) -- (2,0) node[midway, above]{ \scriptsize $l+q$};
		\draw[snake=coil,segment length=4pt] (-5.5,3) -- (-2,0) node[midway, sloped,above] { \scriptsize $\hspace*{.01in}\varepsilon_q,\, q$};
		\draw [snake=coil,segment length=4pt] (-5.5,-3) -- (-2,0) node[midway, sloped,below] { \scriptsize $\hspace*{.2in}\varepsilon_l,\, l$};
		\draw[thick] (5.5,3) -- (2,0) node[midway, above] { \scriptsize $\hspace*{.01in}k_1$};
		\draw [thick] (5.5,-3) -- (2,0) node[midway, below] { \scriptsize $\hspace*{.1in}k_2$};
		\draw (0.7,-4.5) node { \footnotesize (b)};
		\end{tikzpicture}
		\hspace{0.5in}
		\begin{tikzpicture}[scale=.30]
		\draw[thick](2,0) -- (5,3)node[midway, above] {\scriptsize $k_1$} ;
		\draw[thick](2,0) --(5,-3) node[midway,below]{\scriptsize $k_2$};
		\draw[snake=coil,segment length=4pt] (-1.5,3) -- (2,0) node[midway,sloped,above] { \scriptsize $\varepsilon_q,\, q$};
		\draw [snake=coil,segment length=4pt] (-1.5,-3) -- (2,0) node[midway,sloped,below] { \scriptsize $\varepsilon_l,\, l$};
		\draw (2,-4.5) node { \footnotesize (c)};
		\end{tikzpicture}	
\end{center}
\caption{Diagrams contributing to the two tachyon two massless amplitude.}
\label{Allowed}
\end{figure}
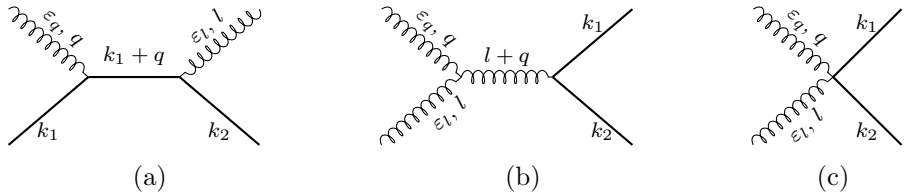

\begin{figure}[]
	\begin{center}
		\begin{tikzpicture}[scale=.30]
		\draw[thick] (-5.5,3) -- (-2,0) node[midway, above]{ \scriptsize $k_1$};
		\draw[thick](2,0) -- (5.5,3)node[sloped, midway, above] {\scriptsize $k_2$} ;
		\draw[snake=coil,segment length=4pt ](-5.5,-3) --(-2,0) node[sloped,midway,below]{\scriptsize $\hspace*{0.1in}\varepsilon_l,\,l$};
		\draw[ultra thick] (-0.6,0.6)--(0.6,-0.6);
		\draw[ultra thick] (-0.6,-0.6)--(0.6,0.6);
		\draw[snake=coil,segment length=4pt] (5.5,-3) -- (2,0) node[sloped, midway,below] { \scriptsize $\hspace*{0.1in} \varepsilon_q,\, q$};
			\draw[snake=coil,segment length=4pt] (-2,0) -- (2,0) node[midway,below]{ \scriptsize $k_1+l$};
		\draw (.5,-3.5) node { \footnotesize (d)};
		\end{tikzpicture}
		\hspace{0.5in}
		\begin{tikzpicture}[scale=.30]
		\draw[ultra thick] (-0.6,0.6)--(0.6,-0.6);
		\draw[ultra thick] (-0.6,-0.6)--(0.6,0.6);
		\draw[thick] (-2,0) -- (2,0) node[midway, above]{ \scriptsize $l+q$};
		\draw[thick](2,0) -- (5.5,3)node[midway, above] {\scriptsize $k_1$} ;
		\draw[thick](2,0) --(5.5,-3) node[near end, left]{\scriptsize $k_2$};
		\draw[snake=coil,segment length=4pt] (-5.5,3) -- (-2,0) node[midway, sloped,above] { \scriptsize $\hspace*{.01in}\varepsilon_q,\, q$};
		\draw [snake=coil,segment length=4pt] (-5.5,-3) -- (-2,0) node[midway, sloped,below] { \scriptsize $\hspace*{.2in}\varepsilon_l,\, l$};
		\draw (1.5,-4.5) node { \footnotesize (e)};
		\end{tikzpicture}
	\end{center}
\caption{Disallowed field theory tree diagrams with a collapsed propagator.}
\label{Notallowed}
\end{figure}
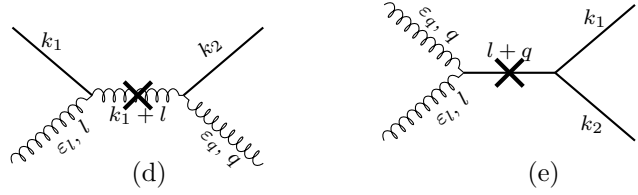

The Feynman diagrams that, in principle, may contribute to the four-point field theory amplitude
are shown in Fig.~\ref{Allowed} and Fig.~\ref{Notallowed}.
They involve all the possible three-point vertices with tachyons and massless states as well as a four-point vertex.  
However, as we will see, the diagrams in Fig.~\ref{Notallowed}
are irrelevant. In fact, the three-point interactions involved in those diagrams, 
do not have a well-defined field theory limit from their corresponding string theory amplitudes, indicating that they are not permitted in the Feynman rules.

We begin with considering the diagrams in Fig.~\ref{Allowed}.
The vertices involved, derived from string theory, were given Sec.~\ref{3ptvertices}.
Based thereon we compute the contribution from each diagram:
\ea{
A^{(a)}_4(k_1,\,k_2, \,q,\,l)&= \varepsilon_{q\mu\nu}\,\varepsilon_{l\rho\sigma}\,\frac{V^{\mu\nu}(k_1,\,q,\,-k_1-q)\,V^{\rho\sigma}(k_1+q,\,l,\,k_2)}{(k_1+q)^2+m^2}+(q\leftrightarrow l)\nonumber\\
&= 2\kappa_D^2\frac{(k_1\epsilon_qk_1)}{k_iq} \big[(k_1\epsilon_lk_1)+(k_1\epsilon_lq)+(q\epsilon_lk_1) +(q\epsilon_lq)\big]+(q\leftrightarrow l)
\label{C.3}
}
where $(q\leftrightarrow)$ denotes the similar contribution from the cross-channel diagram, and we used momentum conservation to replace $k_2=-k_1-l-q$.
By using the on-shell four-point identity $k_1l=k_2q$ and: 
\ea{
\Big[\frac{1}{k_1q}+\frac{1}{k_2q}\Big]=-\frac{(qk_1)(qk_2)}{ql}\, \Big[\frac{2}{(qk_1)(qk_2)} +\frac{1}{(qk_1)^2}+\frac{1}{(qk_2)^2}\Big]=-\frac{ql}{(qk_1)(qk_2)}\label{C.4}
}
we can rewrite Eq. \eqref{C.3} as follows:
\ea{
A^{(a)}_4=&-2\kappa_D^2\frac{(qk_1)(qk_2)}{ql}\Bigg\{(k_1\varepsilon_qk_1)(k_1\varepsilon_lk_1) \Big[\frac{2}{(qk_1)(qk_2)} +\frac{1}{(qk_1)^2}+\frac{1}{(qk_2)^2}\Big]\nonumber\\
&+\Big[\frac{1}{(k_1q)^2}+\frac{1}{(k_2q)(k_1q)}\Big]
(k_1\varepsilon_qk_1) [(k_1\varepsilon_lq)+(q\varepsilon_lk_1)+(q\varepsilon_lq)]\nonumber\\
&+\Big[\frac{1}{(k_2q)^2}+\frac{1}{(k_2q)(k_1q)}\Big]
(k_1\varepsilon_lk_1) [(k_1\varepsilon_ql)+(l\varepsilon_qk_1)+(l\varepsilon_ql)]\Bigg\}\label{C.5}
}

Diagram b gives:
\ea{
A^{(b)}_4=&\,\varepsilon_{q\mu\rho}\,\varepsilon_{l\nu\sigma} \frac{V^{\mu\rho; \nu\sigma;\alpha\beta}(q,\,l,\,-q-l)\,V_{\alpha\beta}(q+l,\,k_1,\,k_2)}{(q+l)^2}
\nn
= &\,
\frac{\kappa_D^2}{2 (ql)}\Big[(\varepsilon_q\varepsilon_l)\,k_1(q-l)\,k_1(q-l)\nonumber\\
& +2[(l\varepsilon_q^t\varepsilon_lk_1)+(l\varepsilon_q\varepsilon_l^tk_1)]k_1(q-l)-2 [(k_1\varepsilon_q^t\varepsilon_lq)+(k_1\varepsilon_q\varepsilon_l^tq)]k_1(q-l)
\nonumber\\
&+4(l\varepsilon_ql)(k_1\varepsilon_lk_1) +4(k_1\varepsilon_qk_1)(q\varepsilon_lq)-4(l\varepsilon_qk_1)(k_1\varepsilon_lq)-4(k_1\varepsilon_ql)(q\varepsilon_lk_1)\Big]
\label{C.6}
}
By use of \eqref{C.4} together with the on-shell identities
$$k_1(k_1+k_2)=k_2(k_1+k_2)=1/2(k_1+k_2)^2=(ql)$$
and
\ea{
(\varepsilon_q\varepsilon_l)\,\frac{k_1(q-l)\,k_1(q-l)}{(ql)}&= -4\frac{(\varepsilon_q\varepsilon_l)(k_1q)(k_2q)}{(ql)}+ (\varepsilon_l\varepsilon_q)(ql)
\nonumber\\
\frac{[(l\varepsilon_q^t\varepsilon_lk_1)+(l\epsilon_q\epsilon_l^tk_1)]k_1(q-l)}{(ql)}&=
2\frac{[(l\varepsilon_q^t\varepsilon_lk_1)+(l\varepsilon_q\varepsilon_l^tk_1)](k_1q)}{(ql)}
+[(l\varepsilon_q^t\varepsilon_lk_1)+(l\varepsilon_q\varepsilon_l^tk_1)]
\nonumber
}
we can rewrite \eqref{C.6} in the following form:
\ea{
A_4^{(b)} =&-2\kappa_D^2\frac{(k_1q)(k_2q)}{ql}\Big[(\varepsilon_q\varepsilon_l)-\frac{[(l\varepsilon_q^t\varepsilon_lk_1)+(l\varepsilon_q\varepsilon_l^tk_1)]}{k_2q}-\frac{[(k_1\varepsilon_q^t\varepsilon_lq)+(k_1\varepsilon_q\varepsilon_l^tq)]}{k_1q}\nonumber\\
&-\frac{(l\varepsilon_ql)(k_1\varepsilon_lk_1) +(k_1\varepsilon_qk_1)(q\varepsilon_lq)-(l\varepsilon_qk_1)(k_1\varepsilon_lq)-(k_1\varepsilon_ql)(q\varepsilon_lk_1)}{(k_1q)(k_2q)}\Big]\nonumber\\
&+\kappa_D^2\Big[\frac{1}{2}(\varepsilon_l\varepsilon_q)(ql)+(l\varepsilon_q^t\varepsilon_lk_1)+(l\varepsilon_q\epsilon_l^tk_1)+(k_1\varepsilon_q^t\varepsilon_lq)+(k_1\varepsilon_q\varepsilon_l^tq)\Big]\label{C.8}
}
The sum of the two diagrams, \eqref{C.4} and \eqref{C.8}, is now seen to give the full four-point amplitude in \eqref{FTA4pt} (upon expanding the double-copy form)
up to contact terms, i.e.:
\ea{
A_4^{(a)}+A_4^{(b)} =&-2\kappa_D^2\frac{(qk_1)(qk_2)}{ql} \Bigg[(\varepsilon_q\varepsilon_l) -\frac{[(l\varepsilon_q^t\varepsilon_lk_1)+(l\varepsilon_q\varepsilon_l^tk_1)]}{k_2q}-\frac{[(k_1\varepsilon_q^t\varepsilon_lq)+(k_1\varepsilon_q\varepsilon_l^tq)]}{k_1q}\nonumber\\
&+\frac{(k_1\varepsilon_qk_1)[(k_1\varepsilon_lk_1)+(k_1\varepsilon_lq)+(q\varepsilon_lk_1)+(q\varepsilon_lq)]}{(qk_1)^2}
\nn
&
+\frac{(k_1\varepsilon_lk_1)[(k_1\varepsilon_qk_1)+(k_1\varepsilon_ql)+(l\varepsilon_qk_1)+(l\varepsilon_ql)]}{(qk_2)^2}\nonumber\\
&+\frac{2(k_1\varepsilon_qk_1)(k_1\varepsilon_lk_1) +(k_1\varepsilon_qk_1)[(k_1\varepsilon_lq)+(q\varepsilon_lk_1)]+ (k_1\varepsilon_lk_1)[(k_1\varepsilon_ql)+(l\varepsilon_qk_1)]}{(qk_1)(qk_2)}\nonumber\\
&-\frac{(l\varepsilon_qk_1)(k_1\varepsilon_lq)+(k_1\varepsilon_ql)(q\varepsilon_lk_1)}{(qk_1)(k_1q)}\Bigg]\nonumber\\
&+\kappa_D^2\,\Big[ \frac{1}{2}(\varepsilon_l\varepsilon_q)(ql)+(l\varepsilon_q^t\varepsilon_lk_1)+(l\varepsilon_q\varepsilon_l^tk_1)+(k_1\varepsilon_q^t\varepsilon_lq)+(k_1\varepsilon_q\varepsilon_l^tq)\Big]\nonumber\\
\equiv& M_4(q,\,l,\,k_1,\,k_2)- \varepsilon_{q\mu\rho} \varepsilon_{l\nu\sigma} V_4^{\mu\rho;\nu\sigma}(q,\,l,\,k_1,\,k_2)\label{C.9}
}
where $M_4$ is the field theory limit of the four-point amplitude computed in App.~\ref{App:4pt} and the contact terms read:
\ea{
\varepsilon_{q\mu\rho} \varepsilon_{l\nu\sigma} V_4^{\mu\rho;\nu\sigma}(q,\,l,\,k_1,\,k_2)=&-
\kappa_D^2\,\Big[\frac{1}{2}(\epsilon_l\epsilon_q)(ql)+(l\epsilon_q^t\epsilon_lk_1)+(l\epsilon_q\epsilon_l^tk_1)+(k_1\epsilon_q^t\epsilon_lq)+(k_1\epsilon_q\epsilon_l^tq)\nonumber\\
&+ 2(k_1\epsilon_q\epsilon_l^tk_1)+2(k_1\epsilon_q^t\epsilon_lk_1)\Big]\label{C.10}
}  
This four-point contact interaction is fully consistent with the double-soft theorem in \eqref{8.3}, as explained in the main text. To the order in the soft expansion considered, only the last two terms contribute there.

Finally, we discuss the disallowed diagrams in Fig.~\ref{Notallowed}.
They involve the vertices of two massless states and one tachyon, as well as a three tachyon vertex.

As done in Sec.~\ref{3ptvertices}, 
we may try to promote the corresponding three-point string amplitudes 
to field theory vertices by stripping off the polarization tensors and keeping 
particle symmetries. We get thereby the following two vertices, respectively:
\ea{
V^{\mu\nu;\alpha\beta}(p_1,p_2,k)&=2\kappa_D\left(\frac{2}{\alpha'}\right)\big[\eta^{\mu\alpha}-\frac{\alpha'}{2} k^\mu\,k^\alpha\big]\big[\eta^{\nu\beta}-\frac{\alpha'}{2} k^\nu k^\beta\big]
\\
V(k_1,\,k_2,\,k_3)&=2\kappa_D\left(\frac{2}{\alpha'}\right)
}
Both vertices are proportional to the inverse of the string slope.
Therefore, we expect that they decouple
in the  $\alpha'\rightarrow 0$ limit performed by keeping fixed  the $D$-dimensional gravitational coupling constant $\kappa_D$.  Indeed, the diagrams in Fig.~\ref{Notallowed}, at leading order in the string expansion and for external on-shell particles,  do not show any poles due to the propagation of particles going on shell. Instead, they become contact terms, which should be described by the four-point vertex. This is easily seen, for example in one of the channels of the diagram in Fig.~(d), one has: 
\ea{
A_4^{(d)}&=\varepsilon_{l\mu\nu}\varepsilon_{q\rho\sigma} \frac{V^{\mu\nu;\alpha\beta}(l,\,k_1,\,-l-k_1)\,V^{~~~~\rho\sigma}_{\alpha\beta}(-k_2-q,\,q,\,k_2)}{(k_1+l)^2}\nonumber\\
&
=\frac{\alpha'}{4}\Big[1-\frac{\alpha'}{2}\frac{k_1l}{1+\frac{\alpha'}{2} k_1l}\Big]\varepsilon_{l\mu\nu}\varepsilon_{q\rho\sigma}\,V^{\mu\nu;\alpha\beta}(l,\,k_1,\,-l-k_1)\,V^{~~~~\rho\sigma}_{\alpha\beta}(-k_2-q,\,q,\,k_2)
\label{C.12}
}
while for the one in Fig.~(e) one gets:
\ea{
A_4^{(e)}&= \varepsilon_{l\mu\nu}\varepsilon_{q\rho\sigma}\frac{V^{\mu\nu;\rho\sigma}(l,\,q,\,-l-q)\,V(l+q,\,k_1,\,k_2)}{(q+l)^2+m^2}\nonumber\\
&=-\frac{\alpha'}{4} \Big[1+\frac{\alpha'}{2} \frac{ql}{1-\frac{\alpha'}{2} ql}\Big]V^{\mu\nu;\rho\sigma}(l,\,q,\,-l-q)\,V(l+q,\,k_1,\,k_2)\label{C.13}
}
In both cases 
the propagator is, as shown, proportional to $\alpha'$ and thus `shrinks to zero', i.e. a point interaction, in the field theory  limit of the amplitude. 
(Taking the vertices into account, in both cases the leading order term of ${\alpha'}^{0}$ is purely local (contact terms), while the on-shell poles only appear at subleading order in ${\alpha'}$.) Thus, in the field theory limit these contributions do not parametrize proper Feynman diagrams, and should instead be described through the four-point contact vertex.

The four point vertex is ensured to take the form in \eqref{C.10} 
by on-shell gauge invariance of the full amplitude.
Thus, we conclude that \eqref{C.10} collects the full four point interaction of two on-shell scalars and massless closed string states.



\begin{thebibliography}{99}








\bibitem{Shapiro:1975cz}
J. A. Shapiro, 	
\href{http://dx.doi.org/10.1103/PhysRevD.11.2937}{{\rm  PRD}
{\bf 11} 2937 (1975)}:
{\it ``On the Renormalization of Dual Models''}


\bibitem{Ademollo:1975pf} 
 M. Ademollo, A. D'Adda, R. D'Auria, F. Gliozzi, E. Napolitano,  S. Sciuto and 
P. Di Vecchia,
\mbox{\href{http://dx.doi.org/10.1016/0550-3213(75)90491-5}{
{\rm Nucl.Phys. B} {\bf 94} 221 (1975)}}:
{\it ``Soft Dilatons and Scale Renormalization in Dual Theories''}


\bibitem{Bianchi:2014gla} 
M.~Bianchi, S.~He, Y.~t.~Huang and C.~Wen,
\href{http://dx.doi.org/10.1103/PhysRevD.92.065022}{{\rm PRD} {\bf 92}, 065022 (2015)},
\href{http://arxiv.org/abs/1406.5155}{{\ttfamily  arXiv:1406.5155}}:
{\it ``More on Soft Theorems: Trees, Loops and Strings''}


\bibitem{Schwab:2014sla}
B.~U.~W. Schwab, 
  \href{http://dx.doi.org/10.1007/JHEP03(2015)140}{{\rm JHEP} {\bf 1503}
  (2015) 140},
\href{http://arxiv.org/abs/1411.6661}{{\ttfamily  arXiv:1411.6661}}:
{\it ``{A Note on Soft Factors for Closed String Scattering}''}




\bibitem{DiVecchia:2015oba} 
  P.~Di Vecchia, R.~Marotta and M.~Mojaza,
\href{http://dx.doi.org/10.1007/JHEP05(2015)137}{{\rm JHEP}
{\bf 1505},  (2015) 137},
\href{http://arxiv.org/abs/1502.05258}{{\ttfamily  arXiv:1502.05258}}:
{\it ``Soft theorem for the graviton, dilaton and the Kalb-Ramond field in the bosonic string''}
  
  


\bibitem{Guerrieri:2015eea}
A.~L.~Guerrieri,
\href{http://dx.doi.org/10.1393/ncc/i2016-16221-2}{{
Nuovo Cim. C \textbf{39}, no.1, 221 (2016)}},
\href{http://arxiv.org/abs/1507.08829}{{\ttfamily  arXiv:1507.08829}}:
{\it ``Soft behavior of string amplitudes with external massive states''}


\bibitem{DiVecchia:2015srk}
P.~Di~Vecchia, R.~Marotta, and M.~Mojaza, 
\href{http://dx.doi.org/10.1002/prop.201500068}{{\em Fortschr. Phys.} {\bfseries 64}
  (2016) 389},
\href{http://arxiv.org/abs/1511.04921}{{\ttfamily  arXiv:1511.04921}}:
{\it ``{Soft Theorems from String  Theory},''}



\bibitem{Bianchi:1512}
M.~Bianchi, A.~ L.~Guerrieri,  
\href{http://dx.doi.org/10.1016/j.nuclphysb.2016.02.005}{{\rm NPB}
  {\bfseries 905}, (2016) 188},
\href{http://arxiv.org/abs/1512.00803}{{\ttfamily  arXiv:1512.00803}}:
{\em ``On the soft limit of closed string amplitudes with massive states''}

\bibitem{DiVecchia:2016amo}
P. Di Vecchia, R. Marotta, M. Mojaza,	
\href{http://dx.doi.org/10.1007/JHEP06(2016)054}{{\rm JHEP } {\bf 1606},  054 (2016)}, 
\href{http://arxiv.org/abs/1604.03355}{{\ttfamily  arXiv:1604.03355}}:
{\em ``Subsubleading soft theorems of gravitons and dilatons in the bosonic string''}

\bibitem{DiVecchia:2016szw} 
 P.~Di Vecchia, R.~Marotta and M.~Mojaza,
\href{http://dx.doi.org/10.1007/JHEP12(2016)020}{{\rm JHEP} {\bf 1612},  020 (2016)}, 
\href{http://arxiv.org/abs/1610.03481}{{\ttfamily  arXiv:1610.03481}}:
{\em ``Soft behavior of a closed massless state in superstring and universality in the soft behavior of the dilaton''}



\bibitem{Sen:2017xjn}
A.~Sen,
  \href{http://dx.doi.org/10.1007/JHEP06(2017)113}{{\rm JHEP} {\bf 1706}
  (2015) 113},
\href{http://arxiv.org/abs/1702.03934}{{\ttfamily  arXiv:1702.03934}}:
{\em ``Soft Theorems in Superstring Theory''}

\bibitem{DiVecchia:2017gfi}  
  P.~Di Vecchia, R.~Marotta and M.~Mojaza,
\href{http://dx.doi.org/10.1007/JHEP10(2017)017}{{\rm JHEP } {\bf 1710},   (2017) 017}, 
\href{http://arxiv.org/abs/1706.02961}{{\ttfamily  arXiv:1706.02961}}:
{\em ``The B-field soft theorem and its unification with the graviton and dilaton''}



\bibitem{Higuchi:2018vyu}
S.~Higuchi and H.~Kawai,
\href{http://dx.doi.org/10.1016/j.nuclphysb.2018.09.024}{
Nucl. Phys. B \textbf{936}, 400-447 (2018)},
\href{http://arxiv.org/abs/1805.11079}{{\ttfamily  arXiv:1805.11079}}:
{\it ``Universality of soft theorem from locality of soft vertex operators''}

\bibitem{Marotta:2019cip}
R.~Marotta and M.~Verma,
\href{http://dx.doi.org/10.1007/JHEP02(2020)008}{
JHEP \textbf{02}, 008 (2020)}
\href{http://arxiv.org/abs/1911.05099}{\ttfamily arXiv:1911.05099}:
{\it ``Soft Theorems from Compactification''}










\bibitem{Sen:2017nim}
A.~Sen,
\href{http://dx.doi.org/ 10.1007/JHEP11(2017)123}{{\rm JHEP}
{\bf 1711}, (2017) 123},
\href{http://arxiv.org/abs/1703.00024}{{\ttfamily  arXiv:1703.00024}}:
{\em ``Subleading Soft Graviton Theorem for Loop Amplitudes''}


\bibitem{DiVecchia:2018dob}
P.~Vecchia, R.~Marotta and M.~Mojaza,
\href{http://dx.doi.org/10.1007/JHEP01(2019)038}{
JHEP \textbf{01}, 038 (2019)},
\href{http://arXiv.org/abs/1808.04845}{\ttfamily arXiv:1808.04845}:
{\it ``Multiloop Soft Theorem for Gravitons and Dilatons in the Bosonic String,''}

\bibitem{DiVecchia:2019kle}
P.~Di Vecchia, R.~Marotta and M.~Mojaza,
\href{http://dx.doi.org/10.1103/PhysRevD.100.041902}{
Phys. Rev. D \textbf{100}, no.4, 041902 (2019)},
\href{http://arxiv.org/abs/1907.01036}{\ttfamily arXiv:1907.01036}:
{\it ``Multiloop soft theorem of the dilaton in the bosonic string''}




\bibitem{DiVecchia:2015bfa} 
  P.~Di Vecchia, R.~Marotta and M.~Mojaza,
 \href{http://dx.doi.org/10.1007/JHEP12(2015)150}{{\rm JHEP } {\bf 1512},  (2015) 150}, 
\href{http://arxiv.org/abs/1507.00938}{{\ttfamily  arXiv:1507.00938}}:
{\it ``Double-soft behavior for scalars and gluons from string theory''}

\bibitem{Klose:2015xoa}
T. Klose, T. McLoughlin, D. Nandan, J. Plefka, G. Travaglini,
\href{http://dx.doi.org/10.1007/JHEP07(2015)135}{{\rm JHEP}\  {\bf 1507},  135 (2015)}, 
  \href{http://arxiv.org/abs/1504.05558}{{\ttfamily  arXiv:1504.05558}}:
{\em ``Double-Soft Limits of Gluons and Gravitons''}

\bibitem{ArkaniHamed:2008gz}
N.~Arkani-Hamed, F.~Cachazo and J.~Kaplan,
\href{http://dx.doi.org/10.1007/JHEP09(2010)016}{
JHEP \textbf{09}, 016 (2010)},
\href{http://arxiv.org/abs/0808.1446}{\ttfamily arXiv:0808.1446}:
{\it ``What is the Simplest Quantum Field Theory?''}





\bibitem{Wang:2015jna}
Y.~Wang and X.~Yin,
\href{http://dx.doi.org/10.1103/PhysRevD.92.041701}{
Phys. Rev. D \textbf{92}, no.4, 041701 (2015)}
\href{http://arxiv.org/abs/1502.03810}{\ttfamily arXiv:1502.03810}:
{\it ``Constraining Higher Derivative Supergravity with Scattering Amplitudes''}

\bibitem{Green:2019rhz}
M.~B.~Green and C.~Wen,
\href{http://dx.doi.org/10.1007/JHEP06(2019)087}{
JHEP \textbf{06}, 087 (2019)}
\href{http://arxiv.org/abs/1904.13394}{\ttfamily arXiv:1904.13394}:
{\it ``Modular Forms and $SL(2, {\mathbb Z})$-covariance of type IIB superstring theory''}



\bibitem{DiVecchia:2015jaq} 
 P.~Di Vecchia, R.~Marotta, M.~Mojaza and J.~Nohle,
\href{http://dx.doi.org/10.1103/PhysRevD.93.085015}{{\rm PRD} {\bf 93}, 085015 (2016)}, \href{http://arxiv.org/abs/1512.03316}{{\ttfamily   arXiv:1512.03316}}:
{\it ``New soft theorems for the gravity dilaton and the Nambu-Goldstone dilaton at subsubleading order''}


\bibitem{DiVecchia:2017uqn}
P.~Di Vecchia, R.~Marotta, M.~Mojaza
\href{http://dx.doi.org/ 	10.1007/JHEP09(2017)001}{{\rm JHEP}\  {\bf 1709}, (2017) 001},
\href{http://arxiv.org/abs/1705.06175}{{\ttfamily  arXiv:1705.06175}}:
{\em ``Double-soft behavior of the dilaton of spontaneously broken conformal invariance''}
 
\bibitem{Guerrieri:2017ujb}
A.~L.~Guerrieri, Y.-t.~Huang, Z.~Li, C.~Wen  
\href{http://dx.doi.org/10.1007/JHEP12(2017)052}{{\rm JHEP}\  {\bf 1712}, 052 (2017)},
\href{http://arxiv.org/abs/1705.10078}{{\ttfamily  arXiv:1705.10078}}:
{\em ``On the exactness of soft theorems''}





\bibitem{Chakrabarti:2017ltl}
S. Chakrabarti, S. P. Kashyap, B. Sahoo, A. Sen and M.~Verma,	
 \href{http://dx.doi.org/10.1007/JHEP12(2017)150}{{\rm JHEP}  {\bf 1712}, (2017) 150},
\href{http://arxiv.org/abs/1707.06803}{{\ttfamily  arXiv:1707.06803}}:
{\it ``Subleading Soft Theorem for Multiple Soft Gravitons''}

\bibitem{Chakrabarti:2017zmh}
S. Chakrabarti, S. P. Kashyap, B. Sahoo, A. Sen and M.~Verma,	
 \href{http://dx.doi.org/10.1007/JHEP01(2018)090}{{\rm JHEP}  {\bf 2018}, (2018) 090},
\href{http://arxiv.org/abs/1709.07883}{{\ttfamily  arXiv:1709.07883}}:
{\it ``Testing Subleading Multiple Soft Graviton Theorem for CHY Prescription''}




%
\bibitem{Scherk:71}
J. Scherk, 
\href{http://dx.doi.org/10.1016/0550-3213(71)90227-6}{Nucl. Phys. {\bf B} 31 (1971) 222}:
{\it ``Zero slope limit of the dual resonance model''}


\bibitem{Witten:2013pra}
E.~Witten,
 \href{http://dx.doi.org/10.1007/JHEP04(2015)055}{
JHEP \textbf{04}, 055 (2015)}
\href{http://arxiv.org/abs/1307.5124}{\ttfamily arXiv:1307.5124}:
{\it ``The Feynman $i \epsilon$ in String Theory''}  {and references therein.}


\bibitem{Tseytlin:2000mt}
A.~A.~Tseytlin,
 \href{http://dx.doi.org/10.1063/1.1376129}{
J. Math. Phys. \textbf{42}, 2854-2871 (2001)}
\href{http://arxiv.org/abs/hep-th/0011033}{\ttfamily hep-th/0011033}:
{\it ``Sigma model approach to string theory effective actions with tachyons''}



\bibitem{KoemansCollado:2019ggb}
A.~Koemans Collado, P.~Di Vecchia and R.~Russo,
 \href{http://dx.doi.org/10.1103/PhysRevD.100.066028}{
Phys. Rev. D \textbf{100}, no.6, 066028 (2019)},
\href{http://arxiv.org/abs/1904.02667}{\ttfamily arXiv:1904.02667}:
{\it ``Revisiting the second post-Minkowskian eikonal and the dynamics of binary black holes''}






   
\end{thebibliography}
\end{document}